\documentclass[12pt,preprint]{aastex}

\begin{document}

\title{An X-ray census of young stars in the Chamaeleon I North cloud}

\author{Eric D. Feigelson\altaffilmark{1,2} \and
Warrick A. Lawson\altaffilmark{2}}

\altaffiltext{1}{Department of Astronomy \& Astrophysics,
Pennsylvania State University, University Park PA 16802}

\altaffiltext{2}{School of Physical, Environmental \& Mathematical
Sciences, University of New South Wales, Australian Defence Force
Academy, Canberra ACT 2600, Australia}

\slugcomment{Submitted to the Astrophysical Journal}

\begin{abstract}

Sensitive X-ray imaging surveys provide a new and effective tool to
establish the census of pre-main sequence (PMS) stars in nearby young
stellar clusters. We report here a deep $Chandra$ $X-ray$ $Observatory$
observation of PMS stars in the Chamaeleon I North cloud, achieving a
limiting luminosity of $\log L_t \simeq 27$ erg s$^{-1}$ ($0.5-8$ keV
band) in a $0.8 \times 0.8$ pc region.  Of the 107 X-ray sources, 37
are associated with Galactic stars of which 27 are previously
recognized cloud members. These include three PMS brown dwarfs; the
protostellar brown dwarf ISO 192 has a particularly high level of
magnetic activity. Followup optical photometry and spectroscopy
establishes that 9-10 of the $Chandra$ sources are probably
magnetically active background stars. Several previously proposed cloud
members are also inferred to be interlopers due to the absence of X-ray
emission at the level expected from the $\log L_t - K$ correlation.
No new X-ray discovered stars were confidently found despite the high
sensitivity.

From these findings, we argue that the sample of 27 PMS cloud
members in the $Chandra$ field is uncontaminated and complete down
to $K = 12$ or $M \simeq 0.1$ M$_odot$. The initial mass function
(IMF) derived from our sample is deficient in $0.1-0.3$ M$_\odot$
stars compared to the IMF of the rich Orion Nebula Cluster and
other Galactic populations.  We can not discriminate whether this
is due to different star formation processes, mass segregation, or
dynamical ejection of lower mass stars.

\end{abstract}

\keywords{open clusters and associations: individual (Chamaeleon I) -
stars:low-mass, brown dwarfs - stars: luminosity function,
mass function - stars: pre-main sequence - X-rays: stars}

\section{Introduction \label{introduction.sec}}

After many decades of study of the stellar initial mass function
(IMF), recent progress has been made at defining the population of
substellar objects \citep{Kroupa02}.  Brown dwarfs (BDs) are much
more luminous and have hotter surfaces shortly after birth
compared to their later cooler stages as L and T dwarfs.  It has
thus proved possible to count them in young pre-main sequence
(PMS) stellar clusters associated with nearby star formation
region \citep[e.g.,][]{Hillenbrand00, Muench02, Briceno02,
Preibisch03, Comeron03, Wilking04}.  A strong suggestion has
emerged that the very young BD population is lower in lower
density star formation regions (like the Taurus-Auriga clouds) and
higher density regions (like the Orion Nebula).  Such studies are
important for understanding the efficiency of gravitational
collapse at low masses, the uniformity of the IMF under different
cloud conditions, and the possible ejection of lower mass objects
from multiple systems or circumstellar disks.

X-ray astronomical surveys of young stellar clusters has proved to
be an effective complement to optical and near-infrared (O/NIR)
surveys in defining PMS stellar populations.  X-ray emission is
elevated $10^1-10^4$ above old disk levels in virtually all stars
$10^7$ yrs and younger, providing an easy discriminant between PMS
cluster members and most Galactic field stars \citep{Feigelson99}.
X-rays trace magnetic activity, primarily magnetic reconnection
flaring, which is relatively insensitive to the presence or
absence of circumstellar disks. X-ray samples are thus
particularly effective in recovering young stars where the disks
have dissipated, often classified as weak-lined or post T Tauri
stars. The relieves the bias towards disky and accreting PMS stars
that is often present in samples based on O/NIR surveys.  An
additional feature is a well-established empirical correlation
between X-ray and photospheric brightness that links the X-ray
luminosity function (XLF) with the $K$-band luminosity function
(KLF).  The astrophysical origin of this correlation is uncertain
\citep{Feigelson03}.

While previous generations of X-ray telescope had insufficient
sensitivity and resolution to reveal the bulk of lower mass
members of young stellar clusters, the $Chandra$ $X-ray$
$Observatory$ can penetrate deeply into the PMS IMF.
Intermediate-sensitivity $Chandra$ studies had detected the
majority of the stellar population and a modest fraction of the BD
population in the $\rho$ Ophiuchi, IC 348, $\sigma$ Orionis and
Orion Nebula young stellar clusters \citep{Imanishi01,
Preibisch02, Mokler02, Feigelson02}. X-ray studies thus join the
active O/NIR efforts to identify and characterize future BDs that
are contemporaneous with the PMS protostellar and T Tauri
population.

We report here an unusually deep observation of the northern core
of the Chamaeleon I star forming cloud obtained with $Chandra$.
Cha I was the site of the first X-ray discovered BD based on
$ROSAT$ imagery \citep{Neuhauser98} and a considerable number of
candidate BDs have been reported using O/NIR photometric and
spectroscopic techniques \citep{Cambresy98, Tamura98, Oasa99,
Comeron00, Gomez03, LopezMarti04, Luhman04, Comeron04b}. Our
$Chandra$ observation is about 100 times more sensitive than the
$ROSAT$ observations due to a combination of lower detector
background, wider bandwidth and longer exposure time. If the X-ray
emission extends to substellar masses without change in the
typical $L_x/L_{bol}$ ratio, then we expect to detect many of the
young BDs as well as the entire T Tauri population.

The Cha I cloud is particularly well-suited to studies of young
stellar populations.  First, it is one of the nearest active star
formation regions at $d \simeq 160$ pc\footnote{ {\it Hipparcos}
studies show that dust appears in the Chamaeleon region around 150
pc \citep{Knude98} and the mean distance for seven stars in the
cloud is 168(+14,-12) pc \citep{Bertout99}. We adopt a distance of
160 pc.}.  Second, it is relatively isolated from other star
forming clouds so there is little confusion due to older PMS stars
that have drifted into the field. Third, the stellar population is
relatively rich with a total population of $200-300$ members.  The
population associated with the North cloud core has two additional
advantages: the molecular material is confined to a small region
so many members are only lightly obscured; and the stellar cluster
is sufficiently compact that several dozen members can be studied
in a single X-ray image. This molecular core, 296.5-15.7, has 30
M$_\odot$ of gas in a 0.4 pc (8\arcmin) diameter region with peak
column density $\log N_H \simeq 22.3$ cm$^{-2}$ \citep{Mizuno99}.

The stellar population of the Cha I North region has been surveyed
at many spectral bands.  The deepest of these surveys attain
limits of $R \simeq 22$, $I \simeq 20$ \citep{LopezMarti04},
$J=18.1$, $H=17.0$, $K=16.2$ \citep{Oasa99}, $L \simeq 11$
\citep{Kenyon01}, $\simeq 2$ mJy in the $5-8.5$~$\mu$m, $\simeq$
4~mJy in the $12-18$~$\mu$m band \citep{Persi00}, and $\simeq
150$~mJy at 1.3~mm \citep{Reipurth96}.  It has also been examined
for faint H$\alpha$ emitting stars \citep{Hartigan93,
LopezMarti04} and for X-ray emitting stars down to $\log L_x
\simeq 29.0$ erg s$^{-1}$ in the soft $ROSAT$ $0.3-2.4$ keV band
\citep{Feigelson93}.

\section{$Chandra$ observations and analysis \label{Chandra.sec}}

\subsection{Data reduction and source detection}

\begin{figure}[t]
\centering
\includegraphics[scale=3.]{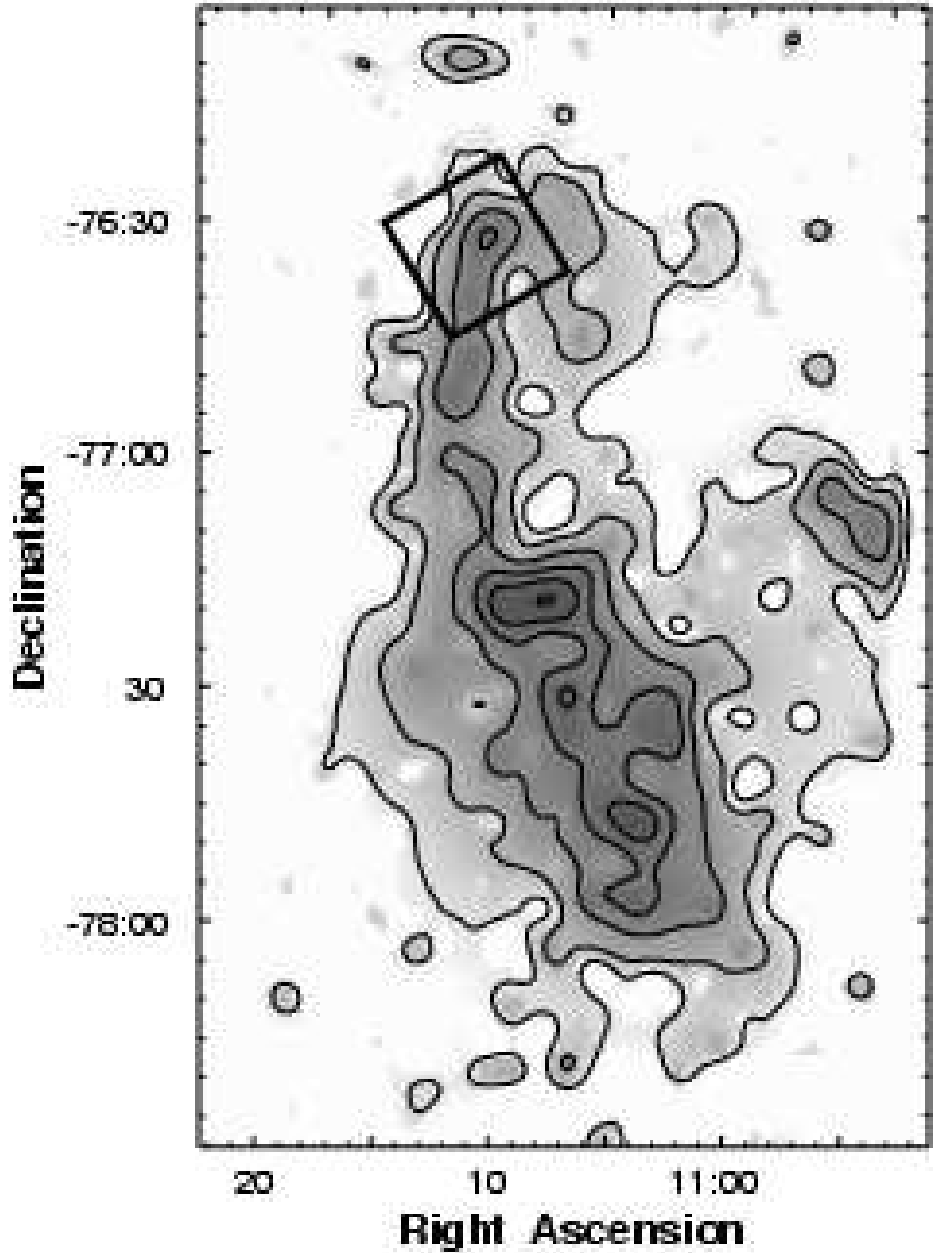}
\caption{Map of the Chamaeleon I cloud showing the $Chandra$ ACIS field.  The map is
in visual magnitudes of absorption derived from $DENIS$ star counts with contours
at $A_V = 1$, 2, 4, 6, 8 and 10 (Cambresy 1997). The box shows the $Chandra$
ACIS field of view.  \label{cloud_abs.fig} }
\end{figure}

A $16\arcmin \times 16\arcmin$ region of the Cha I North cloud was
observed with the imaging array of the Advanced CCD Imaging
Spectrometer (ACIS-I) detector on board the $Chandra$ X-ray
Observatory.  The satellite and instrument are described by
\citet{Weisskopf02}. The detector aimpoint was set at
$(\alpha,\delta) = (11^h10^m0.0^s,-76^\circ35\arcmin00\arcsec)$,
epoch J2000. Figure \ref{cloud_abs.fig} shows the field of view
superposed on the molecular cloud; it subtends a $0.75 \times
0.75$ pc$^2$ region at the cloud. The observation took place on
2.25$-$3.04 July 2001 UT. With 1.3\% of the exposure lost to CCD
readout and 6 s lost to telemetry dropouts, the effective exposure
was 66.3 ks.

The initial stages of data reduction are described in the Appendix
of \citet{Townsley03}. Briefly, we start with the Level 1 events
from the satellite telemetry, correct event energies for charge
transfer inefficiency, and apply a variety of data selection
operations such as ASCA event grades. Several bright sources near
the field center with clear stellar counterparts in the 2MASS
catalog (sources 41, 45, 48, 54, 57 and 62 in Table
\ref{xsrcs.tab}) were used to align the field to the $Hipparcos$
reference frame.  An offset of 0.15\arcsec\/ was applied to the
initial $Chandra$ field position; the individual scatter of these
alignment sources with respect to their 2MASS positions is $\pm\,
0.08$\arcsec.

Candidate sources were located using a wavelet-based detection
algorithm\citep{Freeman02}.  We applied a low threshold ($P = 1
\times 10^{-5}$) so that some spurious sources are found which we
exclude later.  The image was visually examined for possible
additional sources missed by the algorithm; no missing sources or
close double sources were found.  Events for each candidate source
were extracted using the IDL- and CIAO-based script {\it
acis\_extract}\footnote{Description and code for {\it
acis\_extract} are available at
\url{http://www.astro.psu.edu/xray/docs/TARA/ae\_users\_guide.html}.}.
Here, events are extracted in a small region around each source
containing 95\% of the enclosed energy derived from the point
spread function of the telescope at that position. A local
background is defined from a nearby source-free region of each
source and is scaled to the source extraction area.

Unreliable weak candidate sources are now removed.  These include:
sources with $< 3.5$ net (i.e. background-subtracted) extracted
counts; faint sources with median energies above 5 keV that are
probably fluctuations in the background; faint sources with poorly
concentrated events; and faint sources with event arrival times
indicating contamination by cosmic ray afterglows (i.e., several
events appearing in a single pixel within 30 seconds).  Near the
field center (off-axis angle $\theta < 5$\arcmin), source
positions are simple centroids of the extracted events, while
positions for sources far off-axis are obtained from a convolution
of the point spread function with the extracted event positions.
The $acis\_extract$ script also provides position-dependent
telescope-plus-detector effective area $vs.$ energy curves ($arf$
files) and spectral resolution matrices ($rmf$ files) for all
sources.

\begin{figure}[t]
\centering
\includegraphics[width=\textwidth]{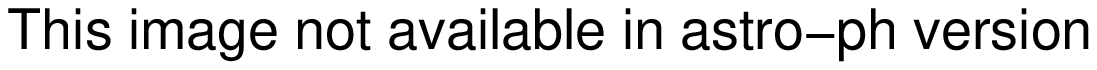}
\caption{$Chandra$ ACIS image of the Cha I North field shown here at
reduced resolution (2\arcsec\/ pixels). Sources are numbered as in
Table \ref{xsrcs.tab}.  Large numbers indicate sources with stellar
counterparts (Table \ref{xprop.tab}), while the others are probably
extragalactic.  \label{acis_img.fig} }
\end{figure}

The resulting 107 $Chandra$ sources and shown in Figure
\ref{acis_img.fig} and their observed properties are given in
Table \ref{xsrcs.tab}.  It gives the running source number, source
position, off-axis angle, background-subtracted extracted counts
(rounded to the nearest count), and cross-reference to the earlier
$ROSAT$ sources designated CHRX \citep{Feigelson93}. Only 17 of
the brightest sources were found with $ROSAT$\footnote{The $ASCA$
satellite, with low spatial resolution but a wide spectral band
similar to $Chandra$'s, detected two sources in our ACIS field
\citep{Ueda01}: 1AXG J110943-7629 ($\log L_x = 30.4$ erg s$^{-1}$)
which blends our sources \#38, 39, 40, 42 and possibly 41; and
1AXG J111011-7635 ($\log L_x = 30.8$ erg s$^{-1}$) which blends
\#53, 56 and 61.  These sources were earlier seen with the
$Einstein$ satellite with similar blending problems
\citep{Feigelson89}.}.

\subsection{Stellar counterparts}

\begin{figure}[t]
\centering
\includegraphics[width=\textwidth]{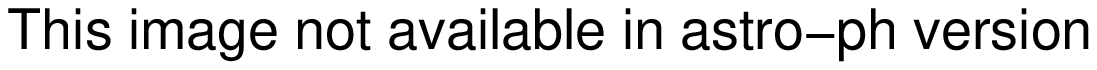}
\caption{Diagram of $Chandra$ stellar sources in the Cha I North
field.  Labeled open circles: Previously known members detected with
$Chandra$.  Crosses: New $Chandra$ sources with stellar counterparts;
source \#16 ($\otimes$) is a tentative new cloud member while the
others are probably background field stars.  The greyscale shows the
cloud cores from the $DENIS$ absorption map (Figure
\ref{cloud_abs.fig}) and the square shows the ACIS field of view.
\label{cloud_srcs.fig}}
\end{figure}

Stellar counterparts are sought within 5\arcsec\/ of the $Chandra$
positions from five databases: the $JHK$ band 2MASS all-sky
catalog, the $IJK$ band DENIS catalog, our $VI$ band CCD images of
the region (\S \ref{opt.sec}), the SIMBAD databases of published
stars, and the list of Cha I cloud members collected from the
literature by \citet{Carpenter02}.  Twenty-seven previously known
cloud members in \citet{Carpenter02} were recovered in the
$Chandra$ observation, nearly all with positional offsets $<
0.5$\arcsec\, as indicated in columns $7-8$ of Table
\ref{xsrcs.tab}). Ten additional X-ray sources associated with
previously unstudied stars are described in \S \ref{new.sec}. We
also discuss in \S \ref{sample.sec} some published candidate
cloud members lying in our $Chandra$ field of view which are not
X-ray detected. Figure \ref{cloud_srcs.fig} shows the stellar
counterparts superposed on the dark cloud.

\subsection{Stellar X-ray spectra and luminosities}

Table \ref{xprop.tab} provide results from subsequent analysis of
the X-ray properties of the stellar sources. The analysis used
$acis\_extract$ version 1.1 for extraction and variability
analysis, and XSPEC 11.2 for spectral modelling. $C_x$ events
were extracted in the polygon containing 95\% of the full point
spread function.  $B_x$ are the background counts scaled to
the extraction region.

The distribution of photon energies were modelled as emission from
a thermal plasma with energy $kT$ based on the emissivities
calculated by \citet{Kaastra00} subject to interstellar
absorption.  The absorption is expressed in equivalent hydrogen
column densities, $\log N_H$ (cm$^{-2}$), assuming solar
metallicities in the intervening gas. (Note, however, that the
recent X-ray absorption study by Vuong et al.\/ 2003 suggests that
dark clouds have metallicities $20-30$\% lower than standard solar
values.) Our modelling is limited by statistical considerations;
weak sources (typically $< 100$ cts) are successfully modelled
with 1-temperature plasmas, stronger sources ($100-1000$ cts)
usually require 2-temperature plasmas, while the strongest sources
($1000-3000$ cts) often require 2- or 3-temperatures with
non-solar elemental abundances. These flux-dependent differences
are unphysical because magnetically active PMS stars, like the
Sun, undoubtedly have continuous distributions of plasma emission
measures over a wide range of temperatures.  The fitted
temperatures represent only the dominant plasma components of the
star during the observation. The presence of non-solar abundances,
particularly involving elements like iron and neon with extreme
first ionization potentials, has been confirmed in high-resolution
X-ray grating studies of stars with strong flare levels
\citep[][and references therein]{Audard03}. We thus caution that
the spectral modeling does not reflect the full range of plasma
properties and, for the fainter sources, may be nearly useless for
interpreting plasma properties due to statistical uncertainties.

However, the broad-band luminosities integrated over the best-fit
model are insensitive to spectral fitting uncertainties and have
roughly $1/\sqrt{C_x}$ errors \citep{Getman02}.  The spectral
model is used here as a nonlinear spline curve through the data.
X-ray luminosities, $L_s$ in the soft $0.5-2$ keV band and $L_t$
in the total $0.5-8$ keV band, are obtained from these fluxes by
multiplying by $4 \pi d^2/0.95$ where $d=160$ pc and 0.95 is the
fraction of photons in the extraction region. These luminosities
are from X-ray emission detected by the $Chandra$ telescope and
represent a lower limit to the emitted luminosities due to
interstellar absorption. The absorption-corrected luminosities
$L_c$ are based on the best-fit spectral model assuming $\log N_H
= 0$.  The accuracy of these values is often low due to
uncertainties in the fitting on nonlinear spectral models to the
observed spectra.  The $L_c$ values should thus be used with
caution.

\subsection{X-ray variability}

Column 5 of Table \ref{xprop.tab} gives a variability indicator
based on the arrival times of the extracted events.  It is coded
to probability of the Kolmogorov-Smirnov test $P_{KS}$ that the
source has no significant variation during the exposure: a =
$P_{KS}>0.05$ (no evidence for variability); b = $0.005 > P_{KS}
> 0.05$ (possibly variable);  c = $P_{KS}<0.005$ (definitely
variable). Note that we expect several `b' values but no `c'
values from a random collection of 107 constant sources.
Lightcurves of prominent flares are shown in Figure
\ref{flares.fig}.

\section{Optical photometry and spectroscopy \label{opt.sec}}

The USNO-B1.0 catalog \citep{Monet03} provides $RI$ photographic
magnitudes and the 2MASS catalog provides $JHK$ magnitudes for
stellar counterparts across the ACIS field of view. There is also
partial coverage of the field within the second release of the
DENIS database at the $iJK$ bands. To augment these surveys, we
obtained $VI$ band CCD images of most of the ACIS field with the
1-m telescope and CCD detector at the South African Astronomical
Observatory (SAAO) during 2002 February. A $15 \arcmin \times
15\arcmin$ field, centered near the $Chandra$ aimpoint, was mapped
with exposure times of 900s and 300s in the $V$ and $I$ bands,
respectively. The SAAO survey covered the Cha I North core, but
missed the extreme corners of the rotated ACIS field (Figure
\ref{acis_img.fig}) where a number of counterparts are located.
Comparison of source magnitudes obtained in the different surveys
(SAAO $I$ $vs.$ DENIS $i$ band magnitudes, and DENIS $vs.$ 2MASS
$JK$ band magnitudes) indicate the various survey magnitudes can
be freely interchanged, with discrepancies mostly comparable to
the photometric errors.  In Table \ref{oprop.tab} we list
USNO-B1.0 $R$, DENIS $i$ or SAAO $I$, and 2MASS $JHK$ magnitudes
for the stellar counterparts.

For three of the brighter candidates, we obtained 2 \AA\,
resolution spectroscopy in the region near H$\alpha$ and covering
the Li I line at 6707.8 \AA\, using the 2.3-m telescope and
dual-beam spectrograph (DBS) at Siding Spring Observatory during
2003 April.  In addition, 4 \AA\, resolution spectra of six of the
candidates were obtained for us by K.\ Luhman, using the Inamori
Magellan Areal Camera and Spectrograph (IMACS) on the Magellan I
telescope.  We discuss the outcomes of our photometric and
spectroscopic analysis of the new $Chandra$ sources in \S
\ref{new.sec}.

\section{Reliability and completeness of the stellar census
\label{complete.sec}}

\subsection{X-ray completeness limit \label{xcomplete.sec}}

Our ability to detect sources in the ACIS field degrades with
off-axis angle (due to broadening of the mirror point spread
function) and with absorption (due to loss of soft X-ray photons).
Feigelson et al.\ (2002, see their \S 2.12) develop formulae for
such detection limits taking into account technical factors such
as optimal extraction radii, telescope vignetting and background
subtraction.  We adopt here a simpler approximation that fits the
lower envelope of sources in Table \ref{xsrcs.tab}.  The limit in
extracted counts as a function of off-axis angle $\theta$ (in
arcmin) is then $C_{lim} \simeq 4$ cts for $\theta < 5\arcmin$ and
$C_{lim} \simeq 4 + (\theta/5)^3$ cts for $5\arcmin < \theta <
11\arcmin$.  The limiting luminosity in the total $0.5-8$ keV band
for a source with a typical PMS spectrum is then $\log L_{t,lim}
\simeq 26.3 + \log C_{lim} + 0.3(\log N_H - 20.0) ~~ {\rm
erg\,s}^{-1}$.  Since cloud absorption is concentrated in the
center of the ACIS field, it is doubtful that both $C_{lim}$ and
$N_H$ are simultaneously high.  For reasonable values of $C_{lim}
\simeq 4-10$ cts and $\log N_H =  21-22$ cm$^{-2}$, the typical
limiting sensitivity across the entire field is then $26.9 < \log
L_{t,lim} < 27.5$ erg s$^{-1}$.

\subsection{Stellar counterpart completeness
\label{counterpartcomplete.sec}}

The O/NIR photometric surveys of Cha I North (\S \ref{opt.sec}) give
counterpart information to limits of $R \approx 20$, $I \approx 18$, $J
\approx 16$, and $H = K \approx 15$. At the distance to Cha I North ($d
= 160$ pc) and adopting a maximum cloud absorption of $A_V \simeq 6-10$
magnitudes, these completeness limits exceed the Zero Age Main Sequence
(ZAMS) for spectral types earlier than M5 ($M > 0.2$ M$_{\odot}$). For
objects free from the regions of high extinction, the sub-stellar
models of \citet{Baraffe98} suggest that all cloud members with $M >
20$ M$_J$ and age $t < 10$ Myr (and still lower masses for younger
ages) will appear as an O/NIR counterpart to any ACIS source.

Unlike earlier satellites where the poor resolution of proportional
counters permitted multiple counterparts, $Chandra$'s excellent point
spread function and satellite alignment give centroids accurate to a
few tenths of an arcsecond on-axis.  Counterpart ambiguities are thus
restricted to multiple systems with component separations less than
$\simeq 100$ AU.

The remaining challenge is then to distinguish stellar from the
dominant extragalactic source population.  This is facilitated by
two tools. First, unresolved extragalactic counterparts of
$Chandra$ sources in our flux range are typically active galactic
nuclei with redshifts $z < 3$ and faint magnitudes, $R \simeq
20-28$ \citep{Hornschemeier01, Barger02}.  They rarely have $R <
19$.  Second, extragalactic sources rarely vary on intra-day
timescales.  Of the 70 non-stellar sources in Table
\ref{xsrcs.tab}, 66 are consistent with constancy, 3 are possibly
variable (consistent with a population of constant sources), and 1
is rapidly variable\footnote{This source \#42 = CXO
110945.1-763022 has 17 events, 16 of which arrived in the final 14
ks of the exposure. Its spectrum is consistent either with a hot
plasma ($kT > 3$ keV) or a powerlaw (photon index $\Gamma \sim 1$)
subject to $\log N_H < 22.0$ cm$^{-2}$ absorption.  This spectrum
could emerge either from a star or an active galactic nucleus.}.
In contrast, 11 of the 37 stellar stellar sources are definitely
variable and 6 are probably variable. Most of the remaining
stellar sources have fewer than 100 photons so only the most
dramatic variations can be seen.

Together, the optical magnitude and the X-ray variability
distributions indicate that little if any erroneous confusion
between extragalactic and stellar sources is present.  The more
subtle distinction between stars associated with the cloud and
background field stars stars is discussed in \S \ref{new.sec} and
\S \ref{sample.sec}.

\section{New stellar counterparts to $Chandra$ sources \label{new.sec}}

We describe here the 10 ACIS sources coincident with O/NIR stars
which have not previously been considered to be candidate members
of the Cha I North star forming cloud. The relevant information
are the photometric magnitudes and colors, our three Siding Spring
Observatory and Magellan spectra (\S \ref{opt.sec}) and the X-ray
properties. The magnitudes and colors alone prove to be a major
constraint.  Most of these stars are too faint to be PMS stars, or
even ZAMS stars, at the distance of the cloud and do not have the
very red colors associated with a low-luminosity young BD.  They
also lie in the outer regions of the ACIS field away from the core
of the young stellar cluster.  We conclude that they are mostly
more distant stars unrelated to the cloud.   In only one case
where a strong emission line is present (\# 16) is it feasible
that the $Chandra$-discovered star may be a cloud member.

Qualitatively, it is not surprising that $Chandra$ detects distant
high-magnetic activity members of the Galactic field main sequence
G, K and early-M population. Studies of solar neighborhood stars
show that $\sim 10$\% of K and early-M disk stars, and $\sim 30$\%
of G disk stars, have X-ray luminosities around $28.0 <\log L_x <
29.5$ erg s$^{-1}$ \citep{Schmitt95, Schmitt97}.  Such stars could
be seen out to distances around $0.5-2$ kpc in our ACIS exposure.
A quantitative study of the background source contamination is
complicated by absorption effects and stellar distributions in the
Galactic disk; this lies beyond the scope of this investigation.

{\bf \#16} ~~ This source lies $\sim 5$\arcmin\/ SW of the cloud core
but suffers considerable soft X-ray absorption with moderate X-ray
brightness. It has $JHK$ colors consistent with a light-to-moderately
reddened ($A_V \simeq 2-3$) early-M or late-K star without $K$-band
excess.  The IMACS spectrum confirms the late-K spectral type and
detects H$\alpha$ in emission suggesting it is a young star.  But the
star is fainter than the ZAMS for its spectral type at the cloud
distance of 160 pc, based on evolutionary models of \citet{Siess00}.
If it is a cloud member, this might indicate the star is being seen in
scattered light, although thus usually occurs only in very heavily
obscured PMS stars. While the preponderance of evidence points to a
background star, we classify this as "Field?".q

{\bf \#22} ~~  This source lies $\sim 6$\arcmin\/ north of the
cloud core and is one of the weakest detected X-ray sources in the
cloud. The $JHK$ colors are consistent with a wide range of
stellar types, from a weakly reddened M3 to a moderately ($A_V
\simeq 5$) G star.  But the star lies below the ZAMS for any of
these possible spectral types.  Showing strong interstellar Ca II
absorption, the IMACS spectrum shows this is a background reddened
G or K field star.

{\bf \#24} ~~ This weak X-ray source lies 18\arcsec\/ WNW of
source \#27 (= CHXR 79) which is $\simeq 40$ times brighter in
the ACIS image. The ACIS spectrum is somewhat unusual, peaking
around 1.3 keV and quickly dropping at both higher and lower
energies probably indicating a high $N_H$ column density.  As with
source \#22, degeneracy in the $JHK$ colors permit a spectral type
between G and early-M, but require that it is a background star
for any spectral type and extinction ($A_V \simeq 6-10$).  IMACS
spectroscopy also indicates a heavily reddened background star,
although of uncertain spectral type.

{\bf \#50} ~~ This weak source lies $\sim 8$\arcmin\/ N of the
cloud core near the edge of the ACIS field. The ACIS spectrum
appears moderately absorbed with best fit column density
equivalent to $A_V \sim 5$, but statistical errors are consistent
with the ambient cloud column density of $A_V \simeq 1-2$. The DBS
and IMACS spectra indicate a weakly-reddened G star with strong Ca
II absorption showing it must lie considerably beyond the cloud.

{\bf \#79} ~~ This weak X-ray source appears about 7\arcmin\/ NE
of the cloud core.  A brief flare may have occurred during the
observation when 10 photons arrived within 1.5 hrs when only $\sim
1$ photon is expected from the remainder of the observation. The
$JHK$ colors are similar to star \#22, and we thus believe it is
another background field star.

{\bf \#84} ~~ This weak source lies about 1\arcmin\/ NE of the
previous new X-ray star \#79.  The spectrum peaks at 1 keV and is
consistent with no absorption, although moderate absorption up to
a few magnitudes in $A_V$ can not be excluded.  The $JHK$ colors
suggest a B or A star with $A_V \simeq 3$, which is confirmed by
DBS and IMACS spectroscopy.  Its faintness requires a distance
considerably further than the cloud.  We note that BA stars
themselves are not thought to be X-ray emitters due to the absence
of an outer convective zone to generate and disrupt magnetic
fields. X-ray emission from such stars is usually attributable to
unseen late-type companions \citep{Daniel02, Stelzer03}.

{\bf \#86} ~~ This source lies 8\arcmin\/ SE of the cloud core and
was detected only by virtue of a dramatic flare during the final 4
hours of the 19 hour observation (Figure \ref{flares.fig}j). The
peak luminosity is $\log L_t \simeq \log L_c = 29.0$ erg s$^{-1}$
and the quiescent luminosity is $< 27.3$ erg s$^{-1}$ if the
distance is 160 pc. The flare spectrum is very hard, exceeding
$Chandra$'s ability to measure the temperature which we estimate
to be $kT \geq 10$ keV. No soft energy absorption is evident. This
is the only X-ray candidate source to show unequivocal
high-amplitude flaring commonly seen in PMS stars.

The stellar counterpart to source \#86 appears as a marginal
detection on photographic sky surveys with $R \simeq 20.5$ and $I
\simeq 18.4$.  It is present at the sensitivity limits of the
2MASS $J$ and $H$ bands and is undetected in their $K$ band. While
source \#86 is thus a candidate low luminosity cloud member, IMACS
spectroscopy indicates a mid-M spectral type, and the distance
modulus indicates it is probably a background flaring field star.

{\bf \#88} ~~ This very faint ACIS source lies $\sim 7$\arcmin\/ W
of the cloud core.  The 2MASS $JHK$ colors are unusual but are
roughly consistent with a lightly reddened ($A_V \simeq 1$) M0
star. The star is too faint at this spectral type to be a cloud
member and we thus classify it as a background field star.

{\bf \#98} ~~ This very weak source lies $\sim 9$\arcmin\/ ENE of
the cloud core in a region of low interstellar obscuration.  The
spectrum is soft with most of the events appearing below 1.2 keV.
We note that such faint and soft sources will be missed in regions
of higher interstellar absorption.  The $JHK$ colors suggest an
unreddened G star and the faint magnitudes require location
considerably beyond the cloud.

{\bf \#99} ~~ This source lies 34\arcsec\/ W of the
weakly-reddened WTT star \#103 = T 50 = Sz 40 about 10\arcmin\/ E
of the cloud core. It has ordinary X-ray properties with spectrum
peaking at 1 keV and a low X-ray brightness.  The DBS spectrum
indicates a G star, consistent with $JHK$ colors with no
reddening.  Here again, the star is too faint to lie at the cloud
distance even if it were on the ZAMS.

\section{Previously known Cha I North stars \label{known.sec}}

Previously identified cloud members from Table \ref{xsrcs.tab} are
briefly described here with a summary of their X-ray properties.
Table \ref{oprop.tab} provides photometric, spectroscopic and
inferred quantities (see table notes for details). Star
designations and cross-identifications with various catalogs are
obtained from \citet{Carpenter02}. Properties on these stars are
obtained as follows:  optical photometry and spectroscopy
\citep{Hartigan93, Huenemoerder94, Alcala95, Lawson96, Saffe03,
LopezMarti04, Luhman04}, pre-$Chandra$ X-ray \citep{Feigelson89,
Feigelson93}, near-IR photometry \citep{Cambresy98, Oasa99,
Persi00, Gomez01, Kenyon01, Carpenter02} and spectroscopy
\citep{Gomez02, Gomez03}. Both optical and near-IR spectroscopy
permit placement of stars on the HR diagram from which masses and
ages can be inferred by comparison with PMS evolutionary tracks.
As discrepancies in reported properties are often present, for
consistency we give spectral types, bolometric luminosities, from
the Magellan spectroscopic study of \citet{Luhman04}, when
available, with masses and ages inferred from \citet{Baraffe98}
evolutionary tracks. Note, however, that other studies often
derive different inferred properties.

For convenience in this section, we use the phrasing: `light',
`moderate' and `heavy' absorption to refer to $A_V<3$, $3<A_V<8$,
and $A_V>8$; `weak', `moderate' and `strong' X-ray emission to
refer to $\log L_c < 28.5$, $28.5 < \log L_c < 29.5$, and $\log
L_c > 29.5$ erg s$^{-1}$; `high', `intermediate' and `low'
bolometric luminosity to refer to $\log L_{bol} \geq 1$, $0.1 <
\log L_{bol} < 1$ L$_\odot$;  and `very young', `young' and
`older' age to refer to $t \leq 1$, $1 < t < 5$, and $t > 5$ Myr.
The nomenclature `WTT' (weak-lined T Tauri) and `CTT' (classical T
Tauri) stars are used here loosely based on a combination of
accretion and disk indicators.

{\bf \#8 CHXR 33 = CHX 13a = C 1-10 = Cam1 65 = ISO 153 = KG 52 =
ChaI 752} ~~ This is a moderately absorbed young M0 WTT star with
intermediate $L_{bol}$ and mass $M \simeq 0.7$ M$\odot$.  It has
strong X-ray emission seen both with the $Einstein$ and $ROSAT$
satellites. The source exhibited a slow 30\% increase over the 19
hour ACIS exposure which could easily arise from evolution in
magnetically confined plasma or in the rotational modulation of
coronal structures.

{\bf \#15 T 37 = Sz 28 = Cam1 68 = ISO 157 = KG 54 = ChaI 709} ~~
This is a moderately absorbed CTT M5 star with strong H$\alpha$
but low $L_{bol}$ indicating an older age. The X-ray emission is
weak.

{\bf \#25 CHXR 35 = Hn 8 = C 1-15 = Cam1 73 = KG 75} ~~  This is a
lightly absorbed M5 $M \simeq 0.2$ M$_\odot$ CTT star with low
$L_{bol}$ and weak X-ray emission.

{\bf \#26 CHXR 37 = Cam1 74 = ISO 185 = KG 77 = RX J1109.4-7627}
~~ This is a high-$L_{bol}$, moderately absorbed, very young K7
WTT star with $M \simeq 0.8$ M$_\odot$. The ACIS count
distribution in the $0.9-1.3$ keV spectral region can be
successfully fit only with weak emission from the iron L-shell
lines, implying a subsolar ([Fe/H] $\simeq 0.3$) abundance of
iron in the heated plasma. This is commonly seen in the X-ray
spectra of magnetically active stars.  The X-ray luminosity is
constant at $\log L_t \simeq 30$ erg s$^{-1}$. This combination of
strong but constant emission is often attributed to heating by
many microflares \citep{Guedel03}.

{\bf \#27 CHXR 79 = Hn 9 = C 1-18 = Cam1 75 = ISO 186 = KG 79} ~~
This is a heavily absorbed, young K7 CTT star with intermediate
$L_{bol}$ and $M \simeq 0.6$ M$_\odot$.  A companion is present
separated by 0.9\arcsec\/ \citep{Brandner96}. The X-ray emission
is strong and constant.

\begin{figure}[t]
\centering
  \begin{minipage}[t]{1.0\textwidth}
    \centering
    \includegraphics[height=0.28\textwidth]{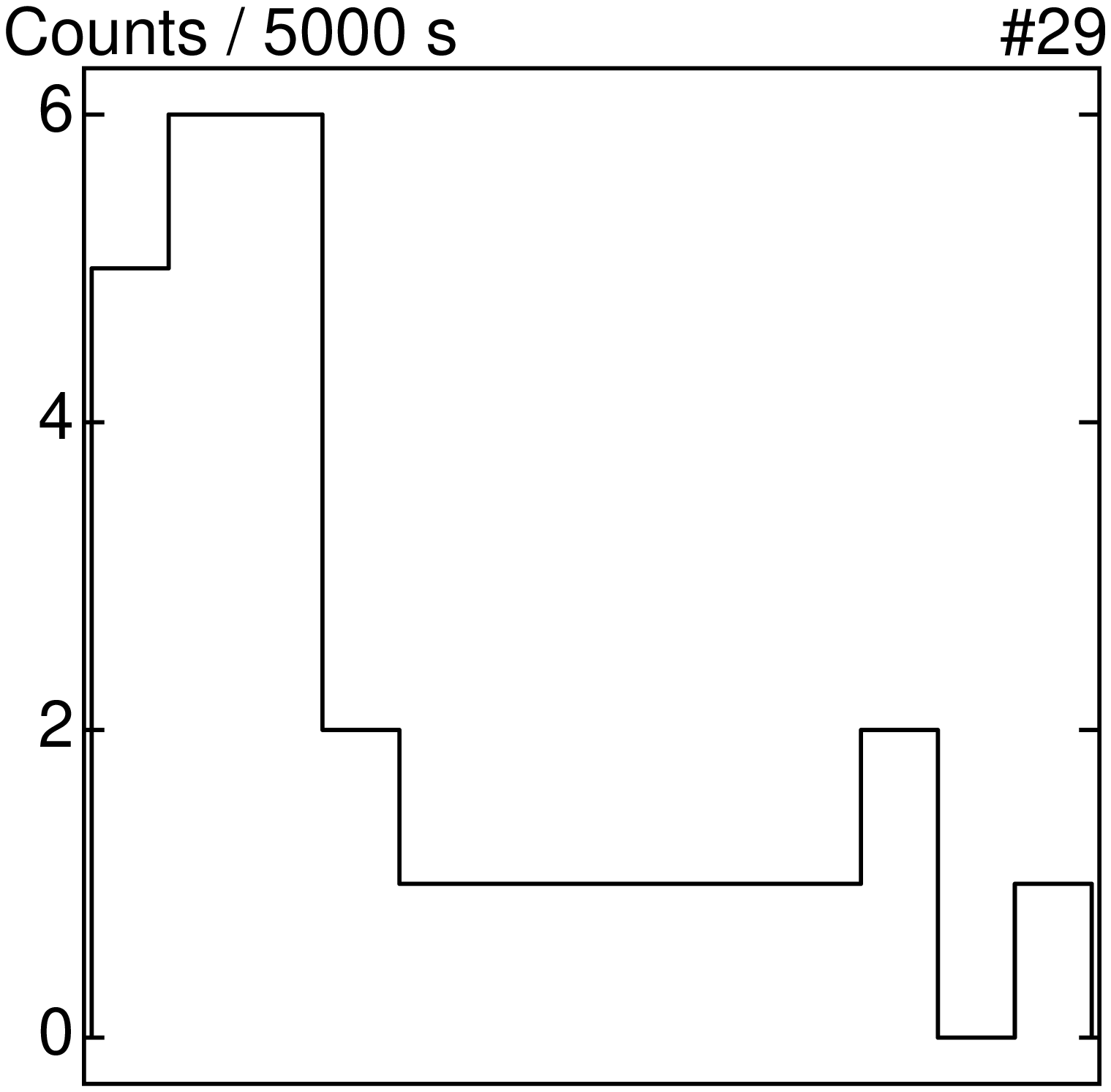}
    \includegraphics[height=0.28\textwidth]{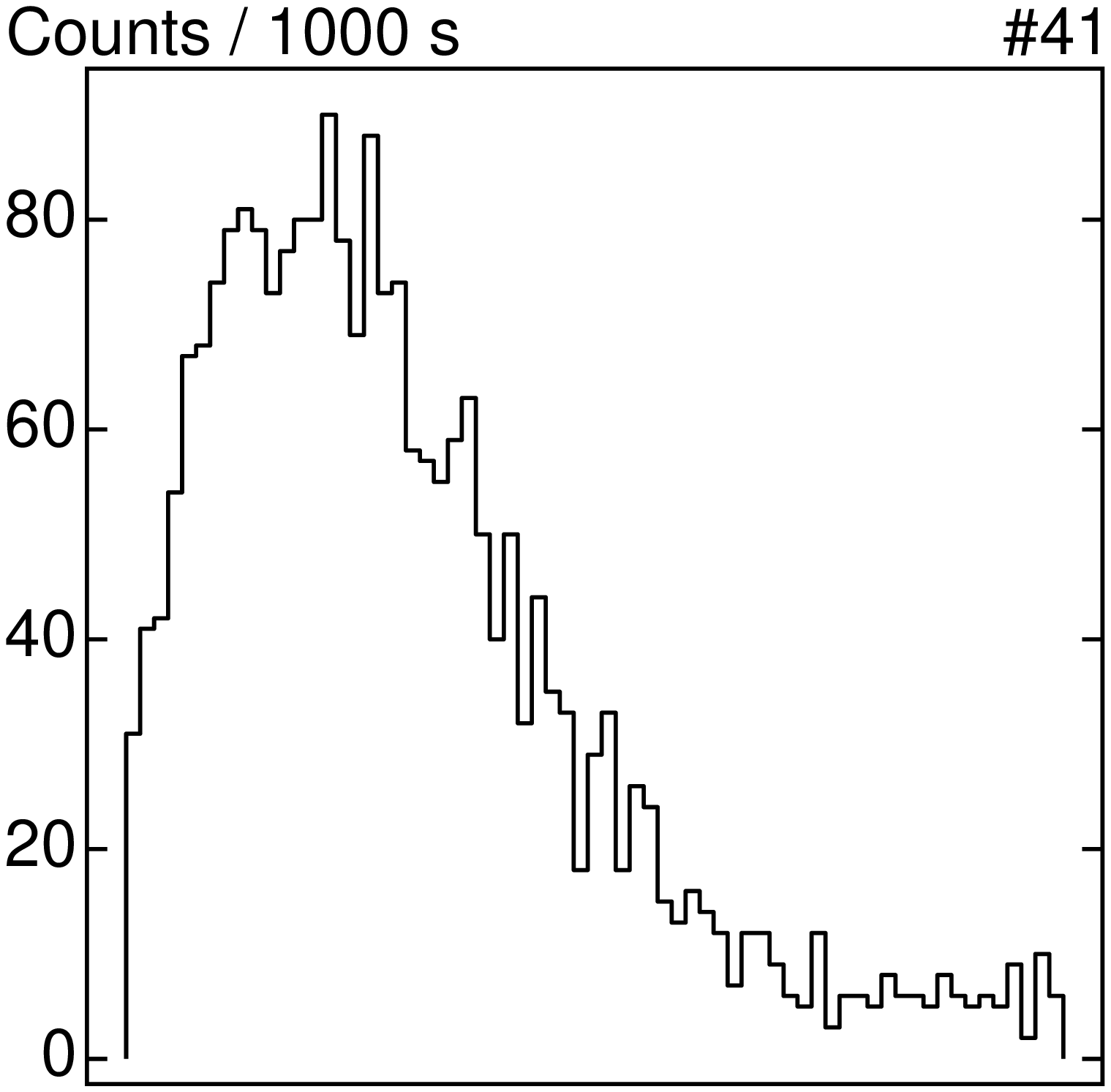}
    \includegraphics[height=0.28\textwidth]{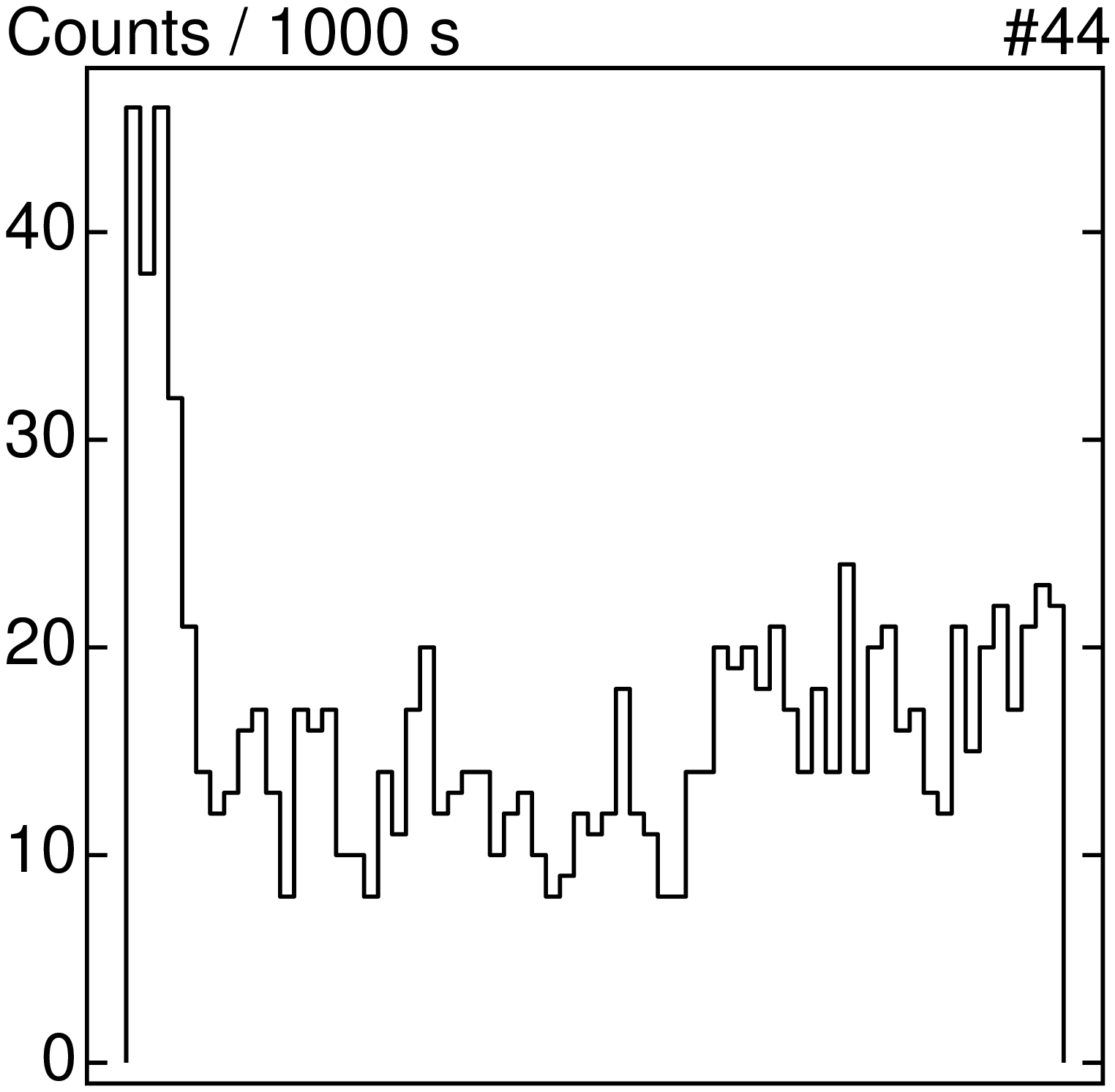}
  \end{minipage} \\ [0.0in]
  \begin{minipage}[t]{1.0\textwidth}
    \centering
    \includegraphics[height=0.28\textwidth]{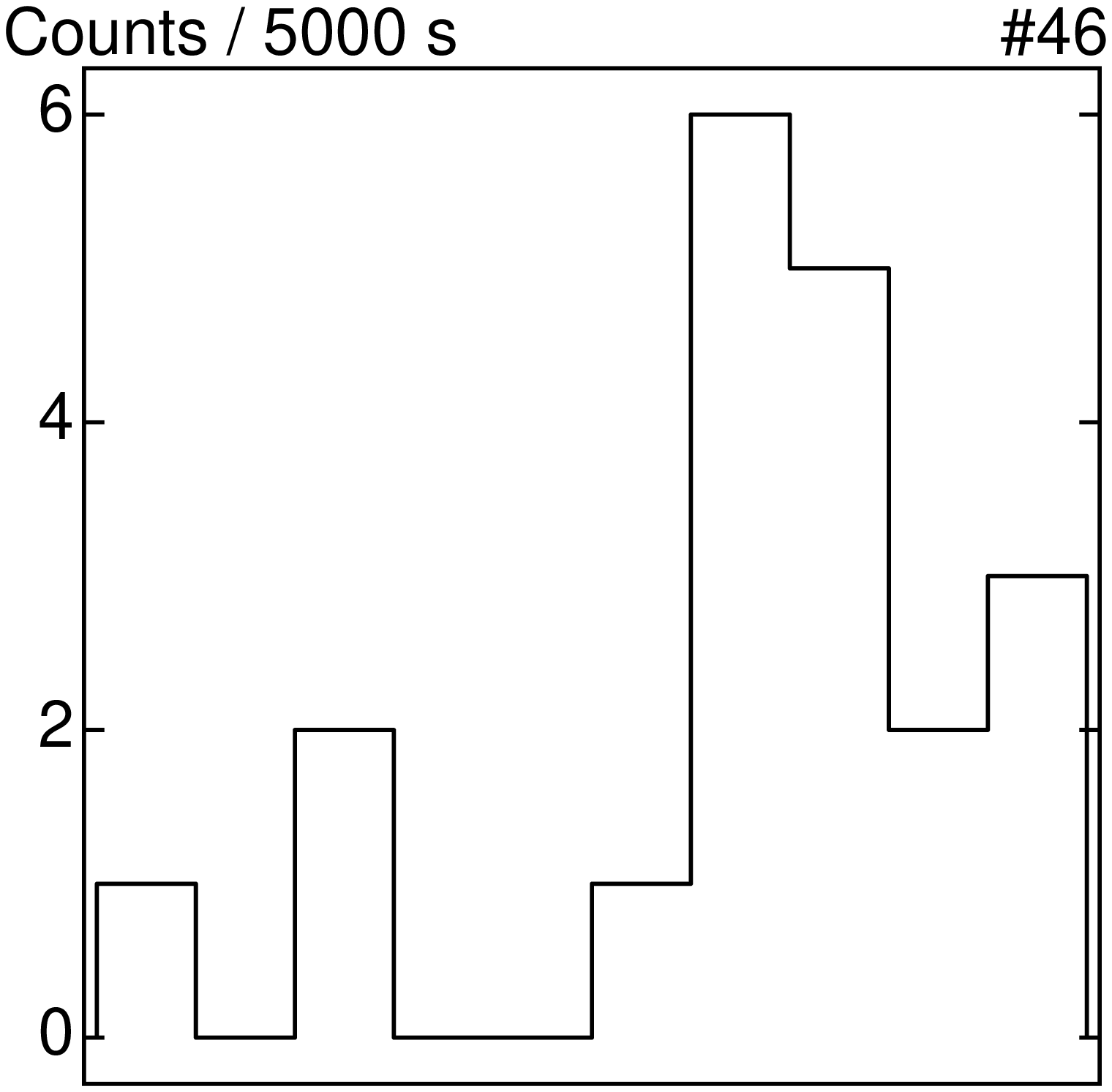}
    \includegraphics[height=0.28\textwidth]{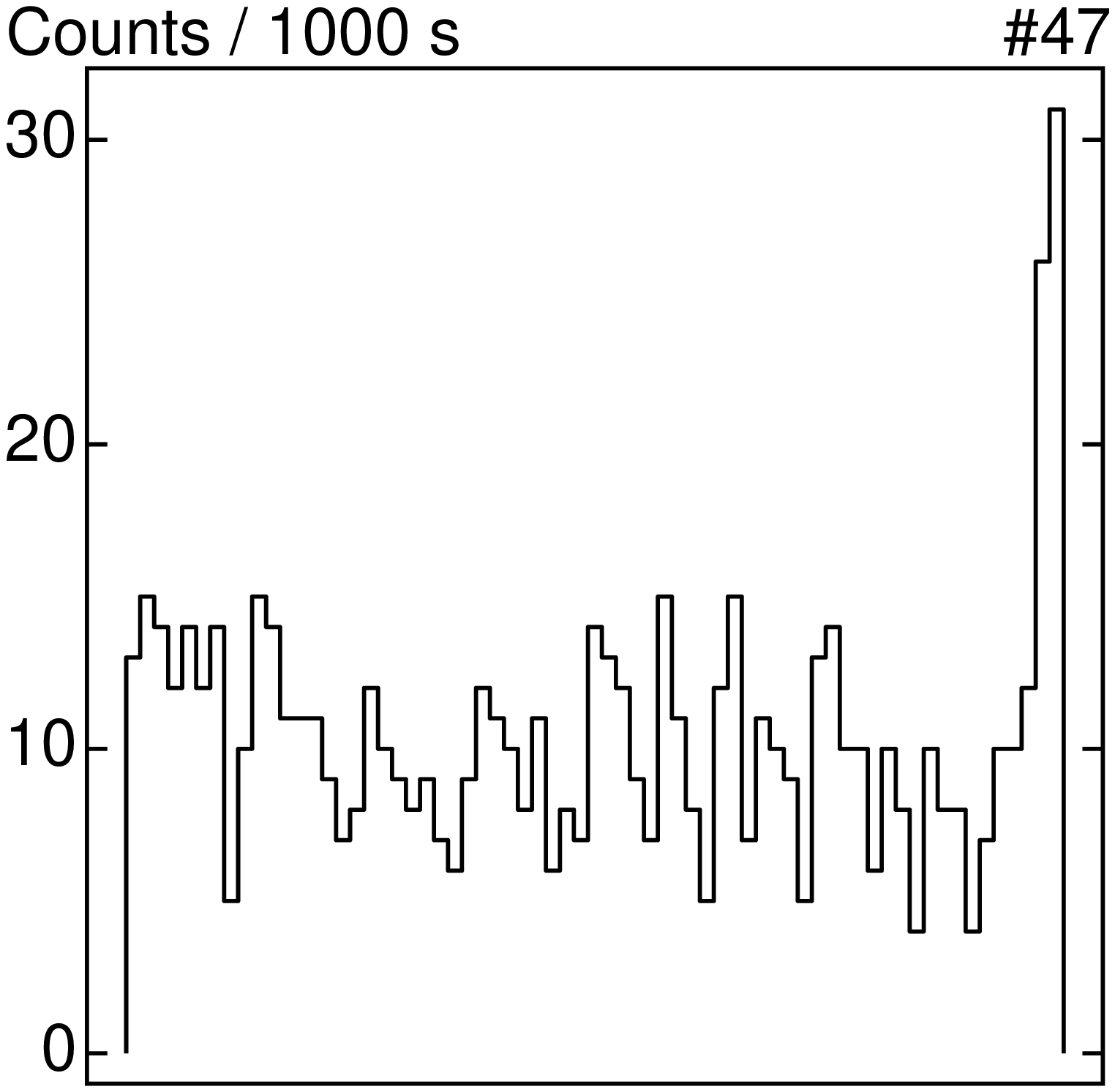}
    \includegraphics[height=0.28\textwidth]{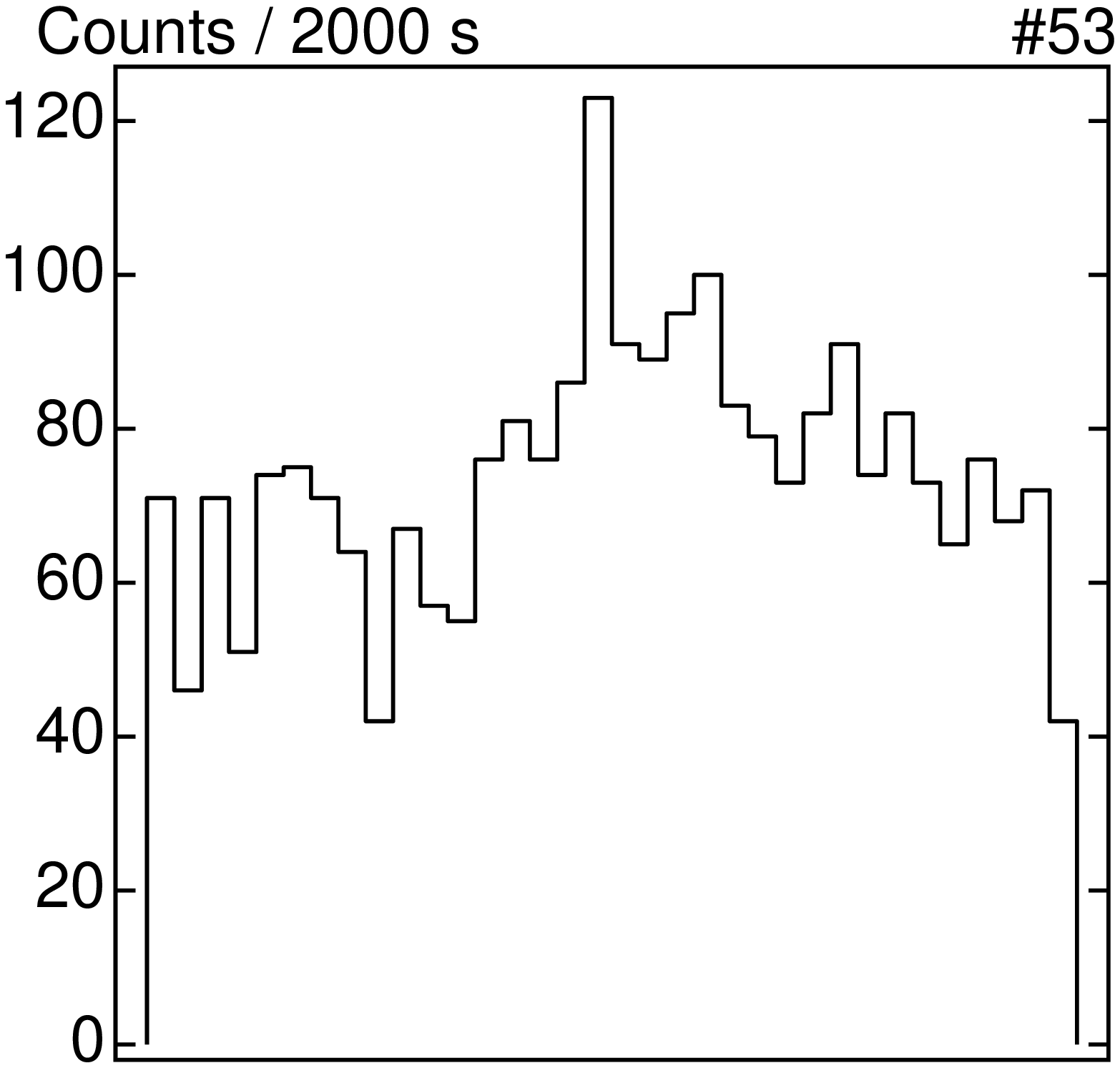}
  \end{minipage} \\ [0.0in]
  \begin{minipage}[t]{1.0\textwidth}
    \centering
    \includegraphics[height=0.28\textwidth]{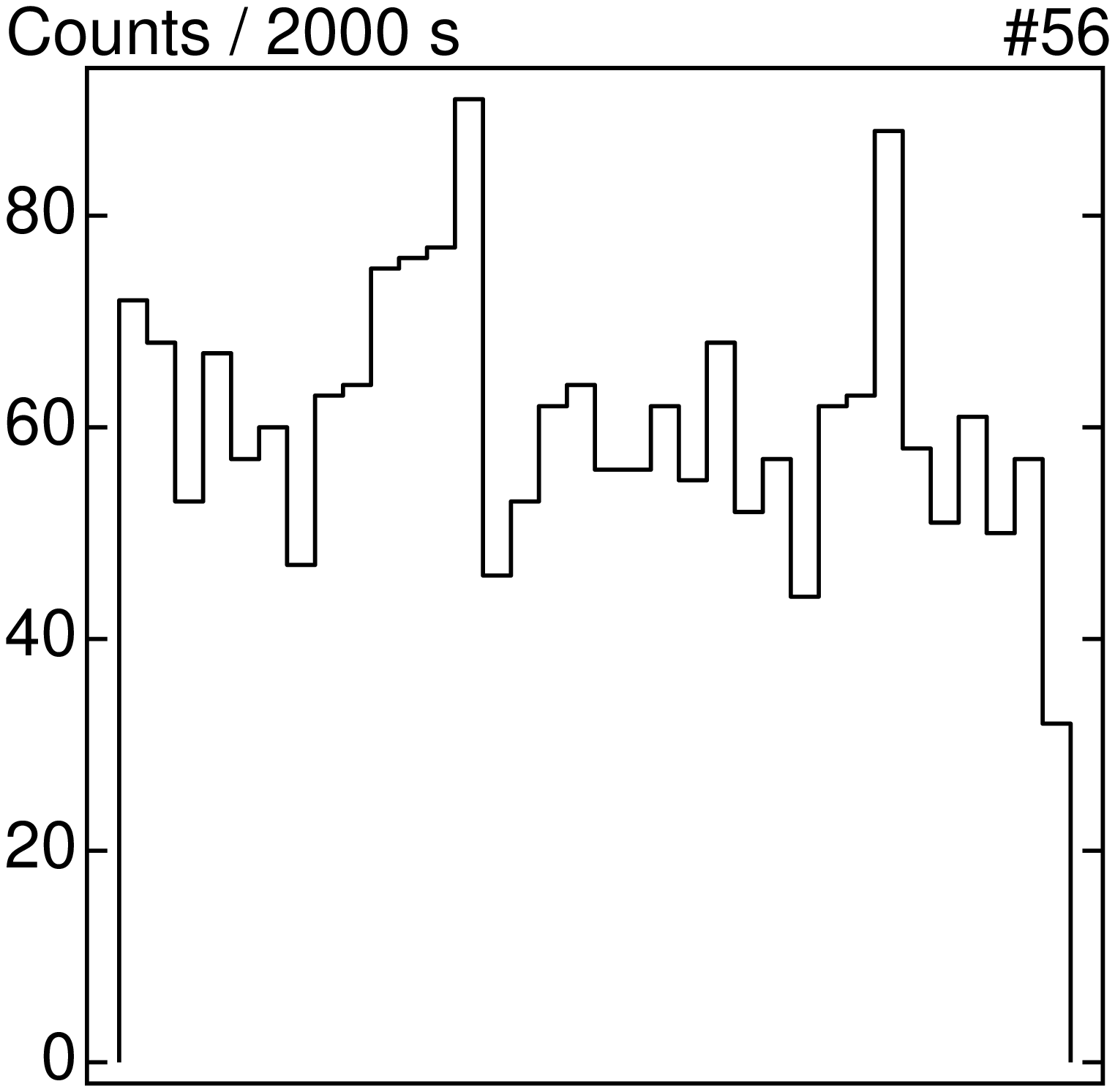}
    \includegraphics[height=0.28\textwidth]{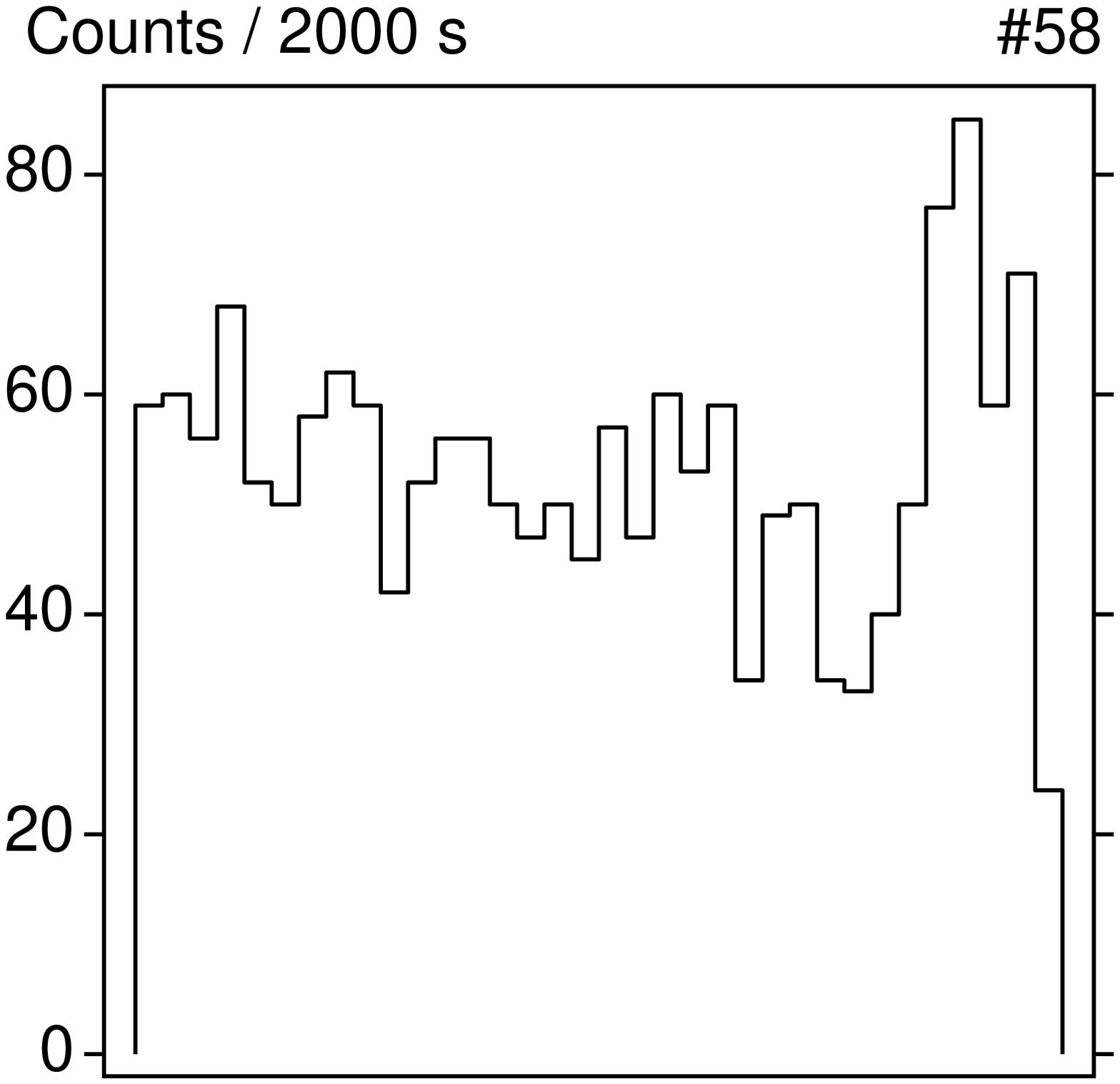}
    \includegraphics[height=0.28\textwidth]{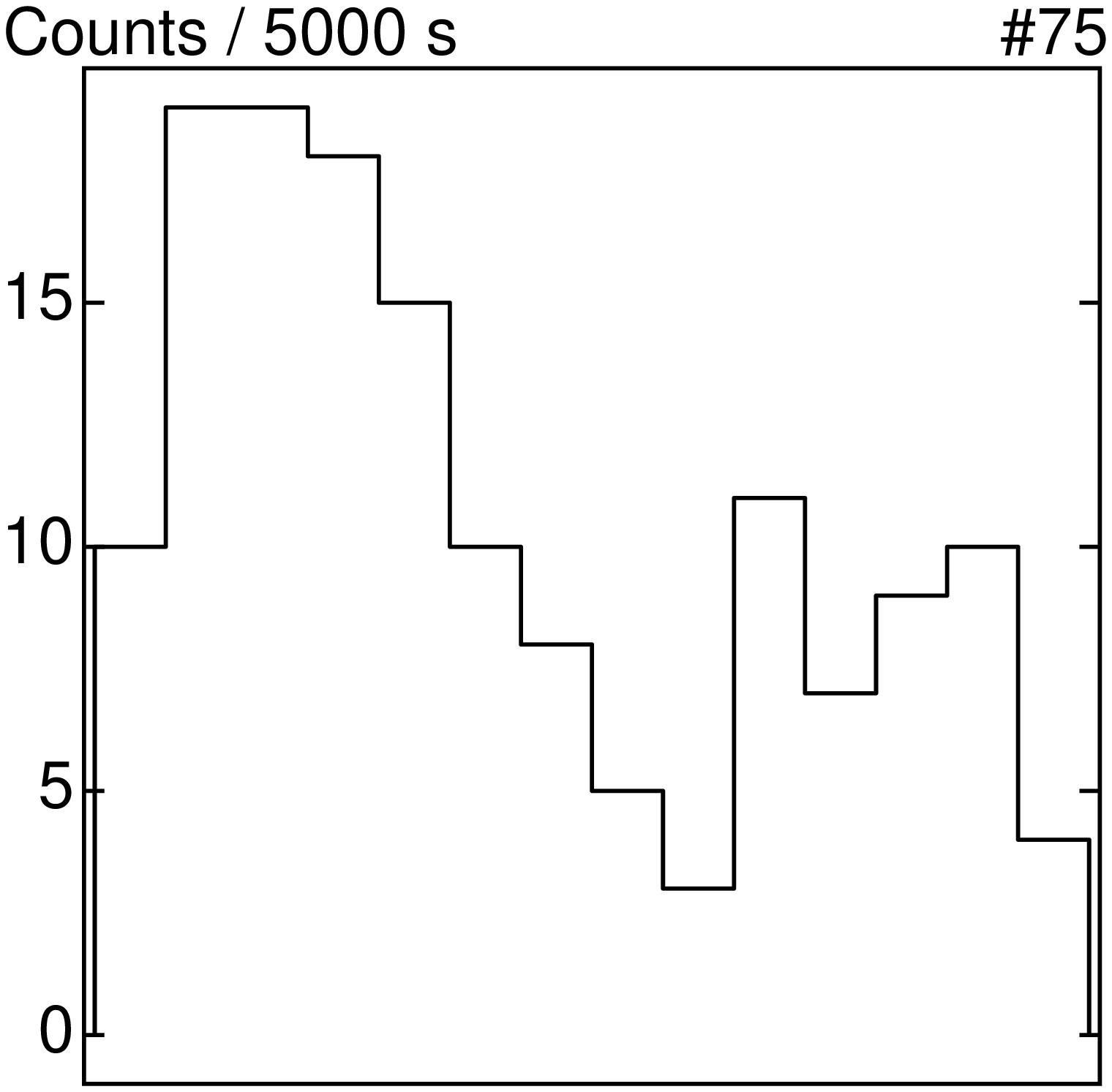}
  \end{minipage} \\ [0.0in]
  \begin{minipage}[t]{1.0\textwidth}
    \centering
    \includegraphics[height=0.28\textwidth]{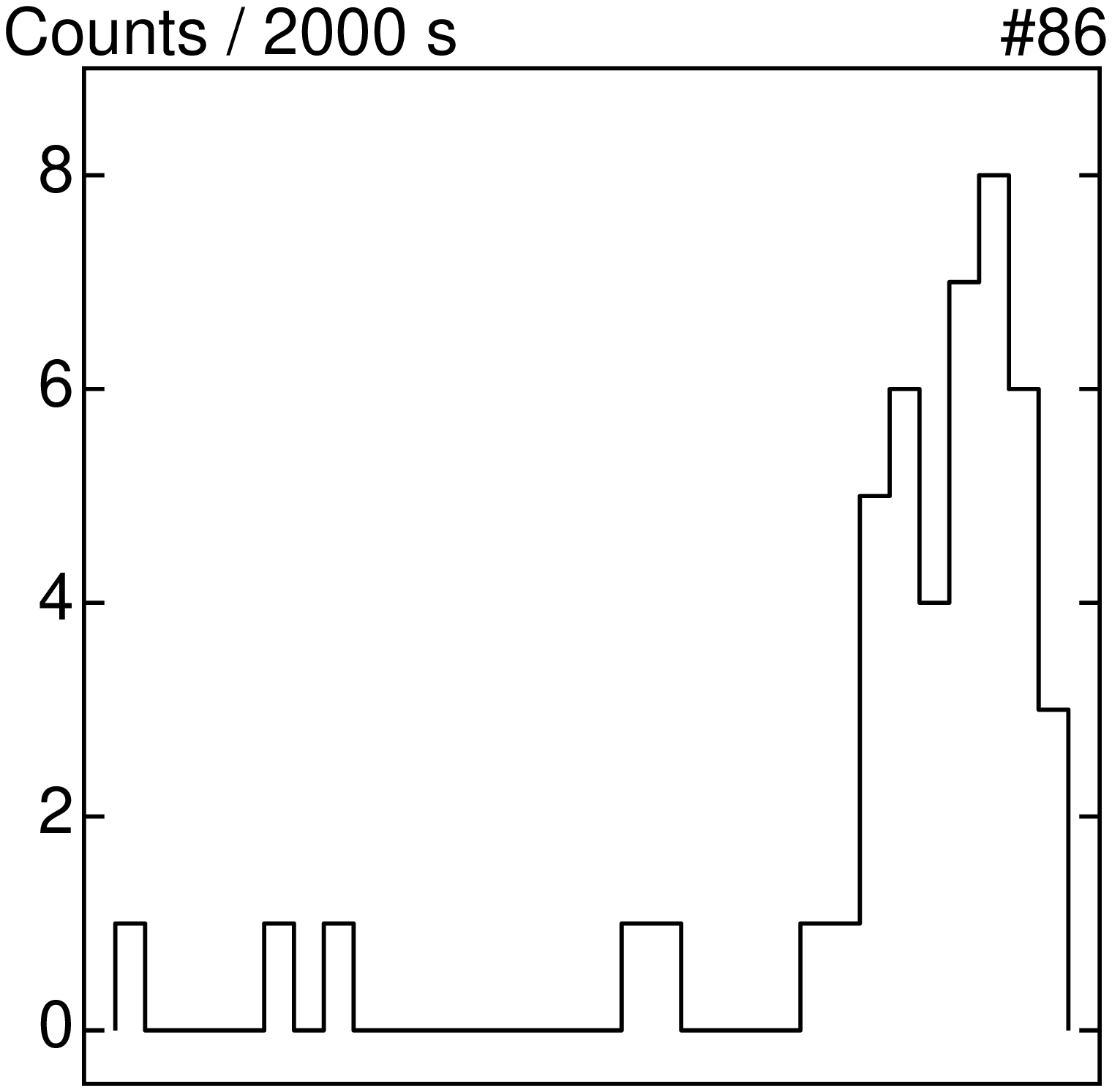}
    \includegraphics[height=0.28\textwidth]{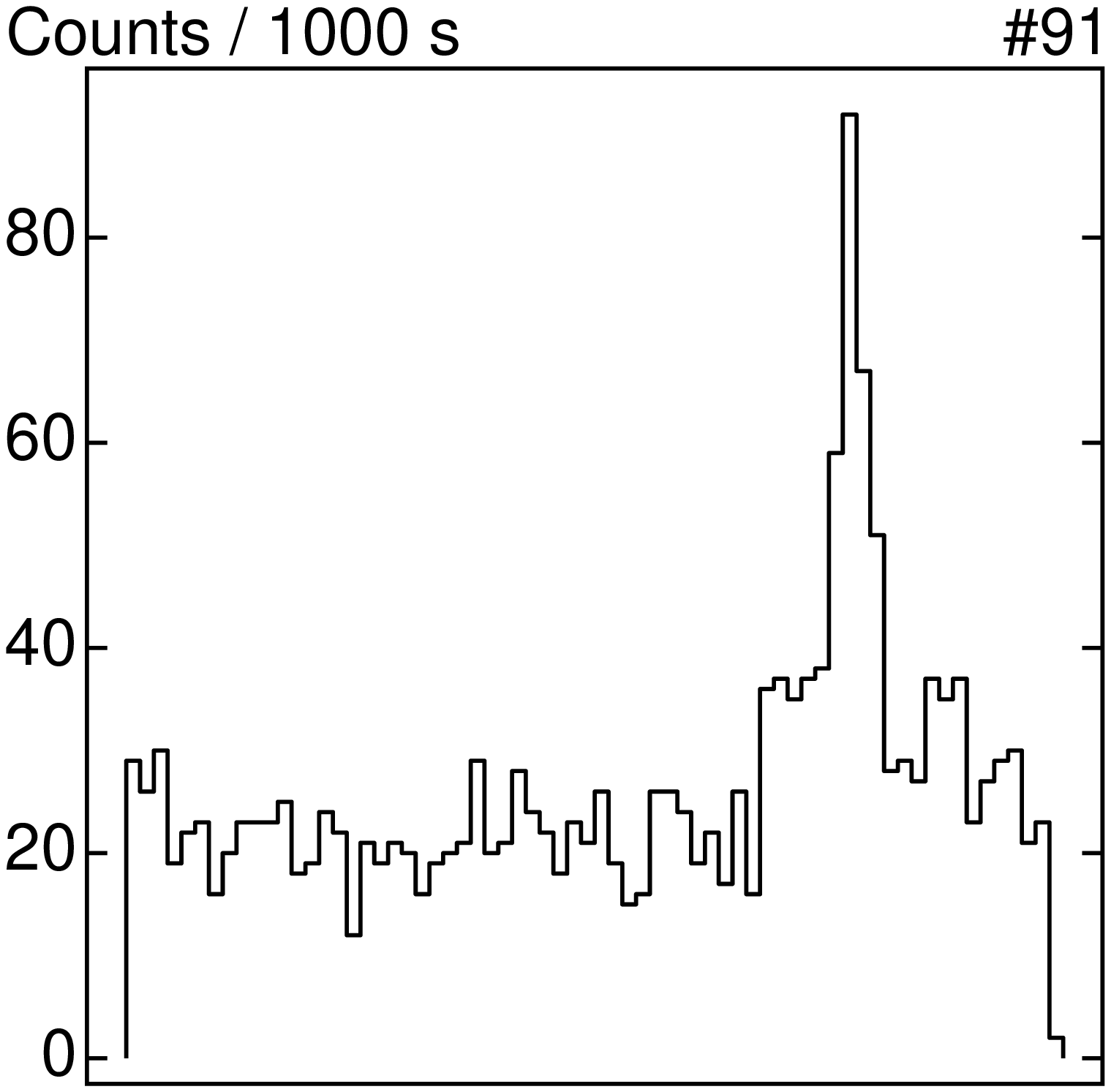}
\caption{$Chandra$ ACIS lightcurves of variable X-ray stars in Cha I
North:  (a) ACIS source \#29 = C 1-6; (b) \#41 = C 1-25; (c) \#44 = T
41 = HD 97300; (d) \#46 = ISO 217; (e) \#47 = T 42 = HM 23; (f) \#53 =
T 44 = WW Cha; (g) \#56 = T 45a = GK 1; (h) \#58 = T 46 = WY Cha; (i)
\#75 = Hn 13; (j) \#86 = background dwarf; (k) \#91 = CHXR 48.  The
abscissa show the photon arrival time from 0 to 66.3 ks.  The ordinates
show counts arrived in bins ranging from 1000 s to 5000 s, as specified
in each panel.  No background has been subtracted and standard
$\sqrt{N}$ errors apply.
\label{flares.fig} }
  \end{minipage} \\
\end{figure}

{\bf \#29 C 1-6 = Cam1 76 = ISO 189 = KG 82 = OTS 10 = ChaI 731}
~~ This is a very young, heavily absorbed, photometrically
variable, M1 CTT star with intermediate $L_{bol}$ and strong
H$\alpha$. The moderate X-ray emission arises entirely from a
flare during the first 3 hours of the exposure with peak $\log L_c
\sim 29.6$ erg s$^{-1}$ (Figure \ref{flares.fig}a).

{\bf \#33 ISO 192 = Cha I Northa2 = Cam2 41 = KG 87 = OTS 15} ~~
This star is an unusual outlier in $JHKL$ color-color diagrams
with the strongest reddening ($A_V \sim 20$) and highest $K$- and
$L$-band excess among Cha I members. This deeply embedded Class I
protostar may be the driving source for the molecular bipolar flow
at the center of the Cha I North cloud core \citep{Persi99}.
Together with its very low bolometric luminosity, these point to a
protostar near or below the substellar boundary \citep{Tamura98}.
Mid-infrared spectroscopy shows strong absorption bands from
silicates and ices characteristic of deeply embedded protostars
\citep{Pontoppidan03, Alexander03}.  Near-IR spectroscopy shows
strong continuum veiling and strong molecular H$_2$ emission
lines. After veiling correction, \citet{Gomez03} derive a spectral
type M6.5 with absorption $A_V \sim 22$, $L_{bol} \simeq 0.06$
L$_{\odot}$, $M \simeq 0.06$ M$_\odot$, and $t \simeq 0.4$ Myr.
Perhaps most surprisingly, the $K-$band continuum has apparently
brightened by over 2 magnitudes during the past several years
\citep{Pontoppidan03}.

The X-ray spectrum is heavily absorbed and, with less than 50
counts, unique spectral parameters can not be obtained.  One
satisfactorily fit has a single $kT = 2$ keV plasma component with
$\log N_H = 22.6$ cm$^{-2}$ ($A_V \sim 25$) giving an intrinsic
absorption-corrected luminosity of $\log L_c \simeq 28.8$ erg
s$^{-1}$.  Other possibilities have absorption ranging up to $\log
N_H \simeq 23$ cm$^{-2}$ ($A_V \sim 80$) with a strong soft plasma
component and $\log L_c$ ranging up to $\simeq 30$ erg s$^{-1}$.
The emission is not dominated by a high-amplitude flare. ISO 192
is thus a relatively strong X-ray emitter compared to most young
BDs.

{\bf \#38 CHXR 40 = Cam1 78 = ISO 198 = KG 92} ~~  This is a
lightly absorbed, very young M1 WTT star with intermediate
$L_{bol}$, $M \simeq 0.6$ M$_\odot$. It may be a wide binary with
a faint 2MASS source 7\arcsec\/ to the south.  The X-ray properties
are similar to those of source \#26: the emission is strong and
constant, best fit with a reduced plasma iron abundance.

{\bf \#41 C 1-25 = Cam1 79 = ISO 199 = KG 93 = OTS 19} ~~  C 1-25
is a heavily absorbed, photometrically variable star with a weak
K-band photometric excess and no emission lines.  Analysis of
near-IR spectra give spectral type M1.5-2, $L_{bol} \simeq 1$
L$_\odot$, $M \simeq 0.2$ M$_\odot$ and $t < 0.1$ My.  It is thus
a very young, low mass WTT star with a remarkably strong X-ray
flare. The X-ray light curve shows a dramatic flare rising steeply
over the first 2 hours of observation, peaking at $\log L_c =
30.7$ erg s$^{-1}$, and then decaying over several hours with
secondary flares to a quiescent level of $\log L_c = 29.7$ erg
s$^{-1}$ (Figure \ref{flares.fig}b). The plasma temperature soared
from 1 keV during the rise phase to $> 10$ keV at the peak,
falling to $\sim 5$ keV during the decay phase and 2 keV at
quiescence.  The column density throughout was steady at $\log N_H
= 22.2$ cm$^{-2}$ ($A_V \simeq 10$).

{\bf \#43 Hn 10E = T 42 = Sz 32 = C 1-24 = Cam1 82 = ISO 204 = KG
97 = OTS 24} ~~  This is an optically bright, moderately absorbed,
very young M3 CTT star with intermediate $L_{bol}$ and mass $M
\simeq 0.2$ M$_\odot$ \citep{Gomez03}.  Its mid-IR spectrum shows
unusually strong silicate emission around 10$\mu$m with no ice
absorption features, possibly indicating a more evolved disk
\citep{Alexander03}. The X-ray luminosity is moderate with a
relatively soft spectrum.

{\bf \#44 T 41 = HD 97300 = CHXR 42 = C 1-11 = Cam1 85 = ISO 211 =
KG 103 = OTS 26 = IRAS 11082-7620} ~~  HD 97300 illuminates the
reflection nebula Ced 112 and is one of the highest mass members
of the Cha I cloud.  With spectral type B9V and mass $M \simeq 2$
M$_\odot$, it has been classified as a Herbig Ae/Be star without
emission lines \citep{The94}.  Its far-IR flux, presumably from a
massive circumstellar disk, is very strong with an $IRAS$ 60$\mu$m
flux of 87 Jy. The primary has a lower mass companion with $\Delta
K \simeq 3$ lying 0.74\arcsec\/ to the NW \citep{Ghez97}. At the
beginning of our X-ray exposure, the emission rapidly ($< 1$ hr)
dropped from $\log L_c \simeq 30.1$ erg s$^{-1}$ to a quiescent
level of $\log L_c \simeq 29.6$ erg s$^{-1}$ which persisted for
the remainder of the observation (Figure \ref{flares.fig}c).  The
flare peak luminosity could have been significantly higher.

The origin of X-ray emission from intermediate mass stars which
have no outer convection zone has been the subject of long debate.
The preponderance of evidence supports an origin in late-type
companions \citep{Feigelson02, Stelzer03}.  In the case of HD
97300, our X-ray position is accurate to $\pm 0.2$\arcsec\/ and
agrees with the primary position to within 0.1\arcsec, so it seems
unlikely that the resolved companion with 0.7\arcsec\/ offset is
the X-ray source.  We suggest that the X-rays arise from a
yet-unresolved third component, although no evidence for
spectroscopic binarity has been found in the primary
\citep{Corporon99}.

{\bf \#46 ISO 217 = KG 106 = GK 29 = ChaI 726} ~~  This is a low
luminosity ($L_{bol} \simeq 0.03$ L$_\odot$) M6 star with estimated
mass $M \simeq 0.07$ M$_\odot$, moderate absorption and a young age.
It is thus a probable proto-brown dwarf.  It also may be a visual
binary. The X-ray source was only by virtue of a flare with
$\log L_{peak} \simeq 29.0$ erg s$^{-1}$; all but one of the 19
events arrived during the last 7 hours of the 18 hours observation
(Figure \ref{flares.fig}d).  The X-ray absorption suggests $A_V
\simeq 30$, but this is poorly constrained by the few counts.

{\bf \#47 T 42 = IRAS 11083-7618 = HM 23 = Sz 32 = C 1-5 = Cam1 86 =
HBC 579 = ISO 223 = KG 109 = OTS 27 = ChaI 753} ~~   HM 23 is one of
the more luminous CTT stars in the Cha I cloud with $L_{bol} = 3.5$
L$_\odot$.  It has a K5 spectrum, estimated mass of 1 M$_\odot$, strong
H$\alpha$ emission, heavy absorption and a massive disk emitting 50 Jy
in the $IRAS$ 60$\mu$m band. The X-ray emission is typically strong for
PMS solar analogs \citep{Feigelson02b}.  A powerful flare with peak
$L_t > 30.5$ erg s$^{-1}$ began during the last 1/2 hour of the
observation (Figure \ref{flares.fig}e).

{\bf \#48 T 43 = Sz 33 = CHXR 41 = CHX 15 = C 1-17 = Cam1 87 = ISO 224
= KG 110} ~~  This is a young, moderately absorbed, photometrically
variable M2 CTT star with an intermediate $L_{bol}$, $M \simeq 0.5$
M$_\odot$, and an infrared disk but relatively weak H$\alpha$ emission.
The X-ray emission, seen earlier with the $Einstein$ and $ROSAT$
satellites, is strong. The luminosity decreased monotonically during
the observation by 40\%, plausibly due either to a slow decay of a
flare or to rotational modulation of magnetic structures.  The X-ray
spectrum shows evidence for reduced iron and enhanced neon in the
emitting plasma.

{\bf \#49 ISO 225 = GK 31} ~~  This is a moderately absorbed,
photometrically variable, very low luminosity ($L_{bol} \simeq 0.013$
L$_\odot$) CTT star in the cloud. The nature of this star has been
difficult to unravel.  The optical spectrum can be modelled as a M2
star from a $M \simeq 0.5$ M$_\odot$ star which is seen only in
reflection \citep{Luhman04}, while the near-IR spectrum can be modelled
as a M5 star with $M \simeq 0.1$ with strong continuum veiling and $A_V
\simeq 5$ \citep{Gomez03}. The X-ray luminosity of $\log L_t \simeq
29.0$ erg s$^{-1}$ is consistent with any PMS star with $M \leq 0.5$
M$_\odot$ \citep{Feigelson03}. The X-ray spectrum indicates an
unusually hot plasma that is heavily absorbed equivalent to $A_V \simeq
30$, considerably higher than inferred from either the optical or
infrared spectrum.  The geometry of obscuring material around this
star, which has the low luminosity of a proto-brown dwarf, may
therefore be unusually complex.

{\bf \#51 C 1-2 = Cam1 88 = ISO 226 = KG 111 = OTS 28} ~~ C 1-2 is
a heavily absorbed CTT star invisible in the optical bands. The
near-infrared spectrum indicates spectral type M1.5, heavy
continuum veiling, $M \simeq 0.2$ M$_\odot$ and a very young age
$<$0.1 Myr \citep{Gomez03}. The X-ray emission is heavily absorbed
($A_V \simeq 20$ with large uncertainty) and intermediate
luminosity.  Though statistically marginal, there may have been a
flare lasting $\sim 6$ hrs during the middle of the observation.

{\bf \#53 T 44 = WW Cha = HM 24 = Sz 34 = C 1-7 = HBC 580 = Cam1
90 = ISO 231 = KG 116 = OTS 29 = CHXR 44} ~~ WW Cha is a luminous,
variable, rapidly rotating ($v$sin$i = 56$ km s$^{-1}$, Franchini
et al.\/ 1988), $M \simeq 0.8-1.0$ M$_\odot$, very young K5 CTT
star with moderate absorption. It is one of the most luminous
stars in the cloud with $L_{bol} \simeq 6$ L$_\odot$. The
mid-IR spectrum shows unusually strong silicate emission around
10$\mu$m with no ice absorption features, possibly indicating a
more evolved disk \citep{Alexander03}.  Its X-ray emission is very
strong, exhibiting a short-lived ($< 1/2$ hour) flare with a peak
flux of $\log 30.6$ erg s$^{-1}$ superposed on a high and more
slowly variable emission (Figure \ref{flares.fig}f).

{\bf \#55 Hn 11 = C 1-4 = Cam1 91 = ISO 232 = KG 117 = OTS 35} ~~
Hn 11 is moderately absorbed, young CTT star with spectral type
K8, intermediate $L_{bol}$ and $M \simeq 0.75$ M$_\odot$. The
X-ray emission is unremarkable with intermediate luminosity,
absorption equivalent to $A_V \simeq 5$, and possible flaring.

{\bf \#56 T 45a = HBC 582 = GK 1 = C 1-9 = HBC 582 = Cam1 92 = ISO
233 = KG 118 = IRAS 11085-7613 = CHXR 45 = CHX 16} ~~ GK 1 is a
lightly absorbed, photometrically variable, $M \simeq 0.7$
M$_\odot$ M0 WTT star with strong X-ray emission seen earlier with
the $Einstein$ and $ROSAT$ satellites.  The X-ray luminosity is
strong, and the lightcurve shows two short ($0.5-2$ hr) flares
superposed on a high quiescent level (Figure \ref{flares.fig}g).

{\bf \#58 T 46 = WY Cha = HM 26 = Sz 36 = C 1-16 = HBC 583 = Cam1
93 = ISO 234 = KG 119 = IRAS 11085-7613 = CHXR 46} ~~ WY Cha is a
lightly absorbed, photometrically variable, young $M \simeq 0.8$
M$_\odot$ K8 CTT star with $L_{bol} \simeq 0.75$ L$_\odot$. The
X-ray emission is strong. The lightcurve shows a flare with peak
$\log L_t> 29.9$ erg s$^{-1}$ superposed on a slower decline
(Figure \ref{flares.fig}h).

{\bf \#61 ISO 237 = C 1-8 = Cam2 48 = KG 121 = OTS 45 = ChaI 760}
~~ ISO 237 is a moderately absorbed, young  K5 CTT star with
estimated mass $M \simeq 0.9$ M$_\odot$ and luminosity $L_{bol}
\simeq 1.3$ L$_\odot$. Its X-ray emission is moderately strong and
constant.

{\bf \#74 T 48 = WZ Cha = HM 28 = Sz 38 = C 1-23 = HBC 585 = Cam1
98 = ISO 258 = CHXR 82} ~~  WZ Cha is a lightly absorbed,
variable, M1 CTT star with $M \simeq 0.7$ M$_\odot$, $L_{bol}
\simeq 0.15$ L$_\odot$ and an unusually strong H$\alpha$ line. Due
to its low luminosity, it has an old inferred age around $t \simeq
10-20$ Myr. This appears to be a case where accretion has endured
for many Myr.  It has weak, soft and unabsorbed X-ray emission.

{\bf \#75 Hn 13 = C 2-5 = Cam1 99 = ISO 259 = ChaI 755} ~~ This is
a lightly absorbed, older M6 CTT star with $L_{bol} \simeq 0.16$
L$_\odot$, $M \simeq 0.08$ M$_\odot$ around the substellar limit.
The X-ray spectrum implies heavier absorption around $A_V \simeq
20$ (with considerable uncertainty) which implies a strong
intrinsic luminosity around $\log L_c \simeq 29.6$ erg s$^{-1}$,
far above the typical level for proto-brown dwarfs.  The
lightcurve shows a 6 hour flare-like variation (Figure
\ref{flares.fig}i).

{\bf \#91 CHXR 48 = E 1-7 = Cam1 101 = ISO 280} ~~ This is lightly
absorbed, photometrically variable M2.5 WTT star with $L_{bol} =
0.25$ L$_\odot$ and $M \simeq 0.4$ M$_\odot$.  It lies on the
periphery of the cloud. The X-ray emission is strong and exhibited
spectacular flaring with both rapid ($\simeq 1$ hr) and slow ($>
6$ hr) components with peak luminosity around $\log L_t \simeq
30.5$ erg s$^{-1}$ (Figure \ref{flares.fig}k).  The X-ray spectrum
implies sub-solar iron abundance in the plasma, as commonly found
in flaring stars.

{\bf \#100 CHXR 84 = Hn 16 = Cam1 105 = E 1-10 = ChaI 742} ~~ This
is a lightly absorbed M5.5 cloud member lying off the cloud core
with estimated mass $M \simeq 0.1$ M$_\odot$ and $L_{bol} \simeq
0.12$ L$_\odot$. With only one measurement of H$\alpha$ with 20
\AA\/ equivalent width and no $K$-band excess, it probably should
be classified as a WTT star.  The X-ray emission is moderate with
a spectral evidence for a reduced iron abundance.

{\bf \#103 T 50 = Sz 40 = E 1-5 = HBC 587 = Cam1 106 = CHXR 85 =
ChaI 757} ~~ This is a very young, photometrically variable M5 CTT
star with $M \simeq 0.1$ M$_\odot$ and $L_{bol} \simeq 0.19$
L$_\odot$. The optical spectrum indicates light absorption of $A_V
\simeq 1$ and it lies in an unobscured region off the cloud core,
but the strong X-ray emission has an unusual spectrum best fit by
a soft plasma suffering heavy absorption equivalent to $A_V \simeq
12$ (with considerable uncertainty).  But with only 60 counts for
analysis, this finding can not be considered definitive.

{\bf \#105 T 51 = Sz 41 = HBC 588 = E 1-9a = Cam1 107 = IRAS
11108-7620 = CHX 20b = CHXR 50} ~~  This is a visually very
bright, unabsorbed, older K3.5 WTT star off the cloud core with
$L_{bol} \simeq 1.1$ L$_\odot$ and $M \simeq 1.2$ M$_\odot$. It is
a visual binary with a fainter companion 1.5\arcsec\/ to the
south-east; the similarly bright star $\simeq 12$\arcsec\/ to the
east is not a cloud member \citep{Ghez97, Walter92}.  It has no
$K$-band excess but nonetheless appears to have a faint outer disk
seen with $IRAS$. The star appears relatively faint during the
$Chandra$ observation with only 50 counts and $\log L_t \simeq
29.0$ erg s$^{-1}$; this is an order of magnitude fainter than
during the earlier $ROSAT$ pointed, $ROSAT$ All-Sky Survey, and
$Einstein$ observations.  However, the source is at the edge
of the ACIS detector and, due to the broad off-axis PSF, many
photons are probably lost during satellite dithering.

\section{Relationships between X-rays and other properties of
PMS stars \label{correl.sec}}

Despite many years of empirical investigation, some fundamental
uncertainties remain in our understanding of the astrophysical
origins of X-ray emission of PMS stars. It is clear that the
X-rays are produced by plasma magnetically confined and heated to
$10^7-10^8$ K.  Violent magnetic reconnection events similar to,
but more powerful than, contemporary solar flares are often seen
in the X-ray lightcurves. But there has been debate regarding the
geometry of the reconnecting magnetic fields: solar-type fields
rooted in the stellar surface; long field lines connecting the
star to the corotation radius of the disk; or fields entirely
associated with the disk \citep{Feigelson99}. It is further
puzzling that X-ray luminosities show strong correlation with the
closely intertwined properties of stellar bolometric luminosity,
size and mass and is not closely linked to stellar rotation as
seen in main sequence stars \citep{Feigelson03, Flaccomio03}. Soft
X-ray absorption can also be compared with optical-infrared
reddening to constraint the gas-to-dust ratio of cloud. We
consider these issues only briefly here because the Cha I North
population is small and our findings are not novel. Stellar
properties are presented in Table \ref{oprop.tab}; for
consistency, most values are adopted from \citet{Luhman04}.

\begin{figure}[t]
  \centering
  \includegraphics[width=0.30\textwidth]{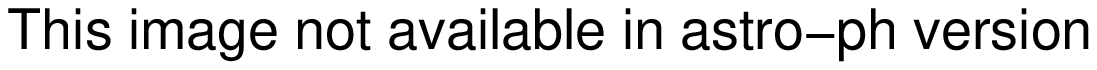}
  \includegraphics[width=0.30\textwidth]{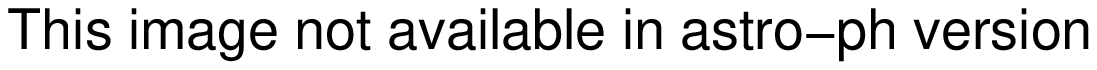}
  \includegraphics[width=0.30\textwidth]{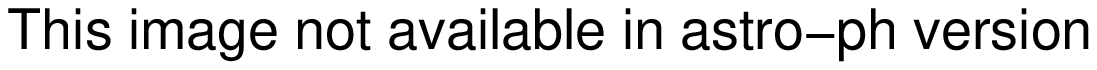}
\caption{Relationships between X-ray luminosity and various
stellar properties.  The four panels compare the observed $0.5-8$
keV luminosities $\log L_t$ (units of erg s$^{-1}$) to: (a) the
observed $K$-band magnitude without correction for absorption or
disk contribution, (b) the stellar luminosity $\log L_{bol}$
(units of erg s$^{-1}$), (c) stellar age (units of Myr), and (d)
stellar mass (units of $M_{\odot}$). In panels (a-d), filled
circles are confirmed cluster members whereas in panel (a) open
circles are unrelated X-ray selected field stars.  In panel (c),
upward arrows indicate cluster members with upper limit ages of 1
Myr.  \label{Lx_prop.fig} }
\end{figure}

Figure \ref{Lx_prop.fig}a shows a broad correlation extending over
$2-3$ orders of magnitude between X-ray luminosity and $K$-band
magnitude\footnote{In this panel, the $Chandra$ sources identified
as field stars are plotted as open circles under the
probably-incorrect assumption they lie at $d=160$ pc. They are
included here to emphasize that the limiting sensitivity of the
$Chandra$ observation lies considerably below the weakest
confirmed cloud members, which is important for evaluation of the
sample completeness (\S \ref{population_sec}).}.  The
$L_x-L_{bol}$ correlation, commonly seen in PMS populations, looks
very similar to this plot.   The similar rough correlation between
$\log L_t$ and mass $M$ (Figure \ref{Lx_prop.fig}b) is expected
from the link between mass and $L_{bol}$ in a roughly coeval PMS
population. A weak anti-correlation between $\log L_t$ and age may
be present (Figure \ref{Lx_prop.fig}c), but this may be an
indirect effect of the below-average bolometric luminosities and
masses of the few older stars in the field.

The outlier with high $L_{bol}$ and $M$ but only average $L_t$ is
HD 97300. The low $L_t/L_{bol}$ ratio of intermediate-mass B and A
PMS stars is often seen in PMS populations, and is usually
attributed to X-ray production by an unresolved magnetically
active low-mass companion rather than the more massive star which
does not have an outer convection zone \citep[e.g.][]{Feigelson02,
Stelzer03}.

Past studies have discussed whether or not the X-ray luminosities
of non-accreting WTT stars are systematically elevated compared to
accreting CTT stars \citep[][and references therein]{Preibisch02,
Flaccomio03}.  The issue is important for interpreting the
astrophysical origin of PMS X-ray emission but is complicated by
sample biases and the $\log L_t - \log L_{bol}$ correlation.
Adopting the classifications given in column 4 of Table
\ref{oprop.tab} and the intrinsic absorption-corrected $0.5-8$ keV
luminosities $\log L_c$, we find mean values of $<\log L_c> = 29.8
\pm 0.2$ erg s$^{-1}$ for 10 cataloged WTT stars and $<\log L_c> =
29.3 \pm 0.2$ erg s$^{-1}$ for 16 cataloged CTT stars. As our
sample is likely complete $M \geq 0.1$ M$_\odot$ (\S
\ref{population_sec}), the reported excess of WTT over CTT X-ray
emission appears to be confirmed.  However, we see no systematic
differences between CTTs and WTT stars in the bivariate $\log L_x
- K$ diagrams or in X-ray variability or spectral. Our results are
thus not conclusive on this issue.

The $N_H$ column densities obtained from X-ray spectral fitting
represent a measurement of the total (i.e. solid, molecular,
atomic and ionized) intervening interstellar gas which can be
compared to the independent optical/IR measurement of reddening by
dust alone \citep{Vuong03}.  The six most heavily absorbed stars
with $>100$ ACIS counts -- CHXR 79, C 1-25, T 42, T 43, T 44, and
ISO 237 ($A_V \simeq 7$) -- have $5 < A_V < 16$. Together these
stars give an interstellar absorption ratio $N_H/A_V \simeq 1
\times 10^{21}$ cm$^{-2}$/mag, lower than the traditional value of
$2 \times 10^{21}$ cm$^{-2}$/mag.  This difference is in the same
direction as the more reliable and well-substantiated value
$N_H/A_V = 1.6 \times 10^{21}$ cm$^{-2}$/mag derived by
\citet{Vuong03} from a larger sample of heavily absorbed PMS stars
in the $\rho$ Ophiuchus cloud.

\section{Discussion: The cloud population \label{population_sec}}

As outlined in \S \ref{introduction.sec}, X-ray surveys provide a
method to obtain the census of a PMS stellar cluster with
different selection effects and different sources of contamination
than encountered in O/NIR surveys. The value of Cha I North is its
proximity, compactness, and unusually intensive study at O/NIR
wavelengths \citep{Cambresy98, Oasa99, Comeron00, Gomez03,
LopezMarti04, Luhman04, Comeron04b}. We combine these with our
$Chandra$ observation giving an very sensitive X-ray survey down
to $\log L_x \simeq 27$ erg s$^{-1}$ to give new insight into the
cloud population and its initial mass function.

\subsection{A complete sample of 27 cloud members
\label{sample.sec}}

We have already argued that our X-ray source list is complete to
$\log L_t \simeq 26.9-27.5$ erg s$^{-1}$ where the value depends
on the individual stars' absorption and location in the ACIS field
(\S \ref{xcomplete.sec}).  This is more sensitive than almost all
previously published studies of young stellar populations which
generally have limiting $\log L_t \geq 28.0$ erg s$^{-1}$. Of the
107 ACIS sources (Table \ref{xsrcs.tab}), 69 are confidently
classified as extragalactic and 37 are confidently classified as
stellar (\S \ref{counterpartcomplete.sec}).  The remaining
ambiguous source (\#42) shows rapid variability resembling a PMS
star but has no O/NIR counterpart.  Of the 37 stellar X-ray
sources, 9 are classified as non-PMS background stars (\S
\ref{new.sec}).  One additional X-ray source (\#16) has a stellar
counterpart that is probably a magnetically active field star. but
could be a cloud member seen with an unusual geometry.

The remaining 27 ACIS sources discussed in \S \ref{known.sec}
constitute the X-ray selected sample of cloud members.  This
sample has no known contaminants and (except for the possible
additions of \#16 and \#42) is complete in X-ray luminosity. There
are no anomalies where a confident bright O/NIR member is
undetected in X-rays.  Based on the correlations between $L_t$,
$L_{bol}$ and $M$ (\S \ref{correl.sec}), we now argue that this
sample is also complete to interesting limits in bolometric
luminosity and mass. Recalling that the O/NIR surveys of the
region are not limited by sensitivity (they can detect all objects
with $M > 20$ M$_J$ with $t < 10$ Myr (\S
\ref{counterpartcomplete.sec}), the issue here is the evaluation
of non-cloud contaminants in various O/NIR samples.

\begin{figure}[t]
   \centering
   \includegraphics[width=0.7\textwidth]{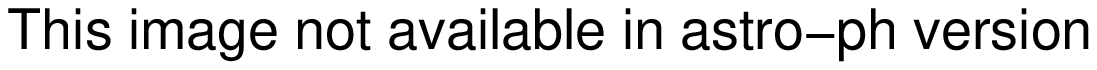}
\caption{Plot of X-ray luminosities $\log L_t$ and $K$ magnitudes for
the complete sample of 27 Chan I North stars (large circles) compared
to a sample of 399 X-ray/infrared stars from the Orion Nebula (small
dots) with $K$ magnitudes adjusted to match the distance of the
Chamaeleon cloud.  The shaded band shows the X-ray sensitivity limit of
the present observation. \label{Lx-K.fig}}
\end{figure}

Figure \ref{Lx-K.fig} compares the $\log L_t - K$ relationship for
the 27 Cha I North stars (previously seen in Figure
\ref{Lx_prop.fig}a) with that seen in the more populous Orion
Nebula Cluster\footnote{The Orion sample is obtained from the
tables of the $Chandra$ Orion Ultradeep Project (COUP) provided by
\citet{Getman04}.  The sample here consists of 399 COUP sources
with counterparts in the $JHKL$ catalog of \citet{Muench02}. The
Orion $K$ magnitudes have been artificially increased by 2.0 to
place them at a distance of 160 pc rather than 450 pc. $\log L_t$
values were obtained using {\it ACIS Extract} package consistent
with our analysis here. This Orion sample should not be viewed as
complete (in particular, there are additional COUP sources with
fainter $K$-band counterparts; M.\ McCaughrean, private
communication), but rather gives a larger PMS population to better
define the correlation and scatter in the $\log L_t - K$
diagram.}.  The Chamaeleon I and Orion samples clearly occupy the
same region in the diagram and show the same correlation. The
shaded band indicates our Cha I North X-ray completeness limit (\S
\ref{xcomplete.sec}).

We use the Orion sample in Figure \ref{Lx-K.fig} to argue that, at
a distance of 160 pc, an X-ray survey complete to $\log L_t \simeq
26.9-27.5$ erg s$^{-1}$ is also complete to $K \simeq 11$ and will
capture the majority of stars with $11 < K < 12$. We are confident
of this conclusion even without knowing the exact nature of the
$\log L_t-K$ relation and its scatter at low luminosities because
no cloud members were found below $\log L_t = 28.2$ erg s$^{-1}$.
We consider this to be important: the ACIS image should have
detected all cloud members between $26.9-27.5 < \log L_t < 28.2$
erg s$^{-1}$.  Half of the extragalactic sources and most of the
field star sources (open circles in Figure \ref{Lx_prop.fig}a)
were found at these low flux levels, so there clearly is no
operational difficulty in detecting such sources. The exception
would be cloud members with extremely high absorption, $\log N_H >
23$ cm$^{-2}$ or $A_V > 100$, such as the Class 0 protostar
Cha-MMS 2 \citep{Reipurth96}. With this caveat, we conclude that
the X-ray complete sample of 27 cloud members represents a
complete and uncontaminated census of cloud members with $K \leq
12$ lying in the ACIS field.

One byproduct of this result is a clarification of the membership
status of several O/NIR stars discussed in the literature. Six
stars -- ISO 154, ISO 164, ISO 165, Ced112 IRS2, ISO 247
\citep{Carpenter02} and ESO-H$\alpha$ 564 \citep{Comeron04b} --
have $10.2<K<11.7$ and are undetected in the ACIS image.  These
can be excluded from cloud membership with considerable
confidence\footnote{The X-ray measurement do not clarify the
membership status of C 1-14, a reddened F0 star with $K=7.8$
because F stars often have soft coronal X-ray spectra which could
be fully absorbed by interstellar gas.}.  Most of these
X-ray-quiet stars have independently been evaluated to be
background stars based on optical spectra by \citet{Luhman04}. The
exception is the M5.5 star ISO 165, a slightly reddened $\sim$M5
star with strong H$\alpha$ but negligible infrared excess
\citep{Kenyon01, LopezMarti04}, which Luhman classifies as a cloud
member but which we believe is a non-member due to the X-ray
non-detection. With only one such disagreement, we find a
gratifying agreement between our X-ray and Luhman's spectroscopic
census of cloud members considering that the selections are based
on very different criteria.

\subsection{The KLF and IMF \label{IMF.sec}}

\begin{figure}[t]
   \centering
   \includegraphics[width=0.45\textwidth]{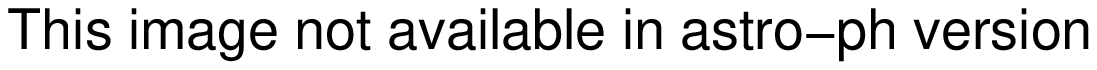}
   \includegraphics[width=0.45\textwidth]{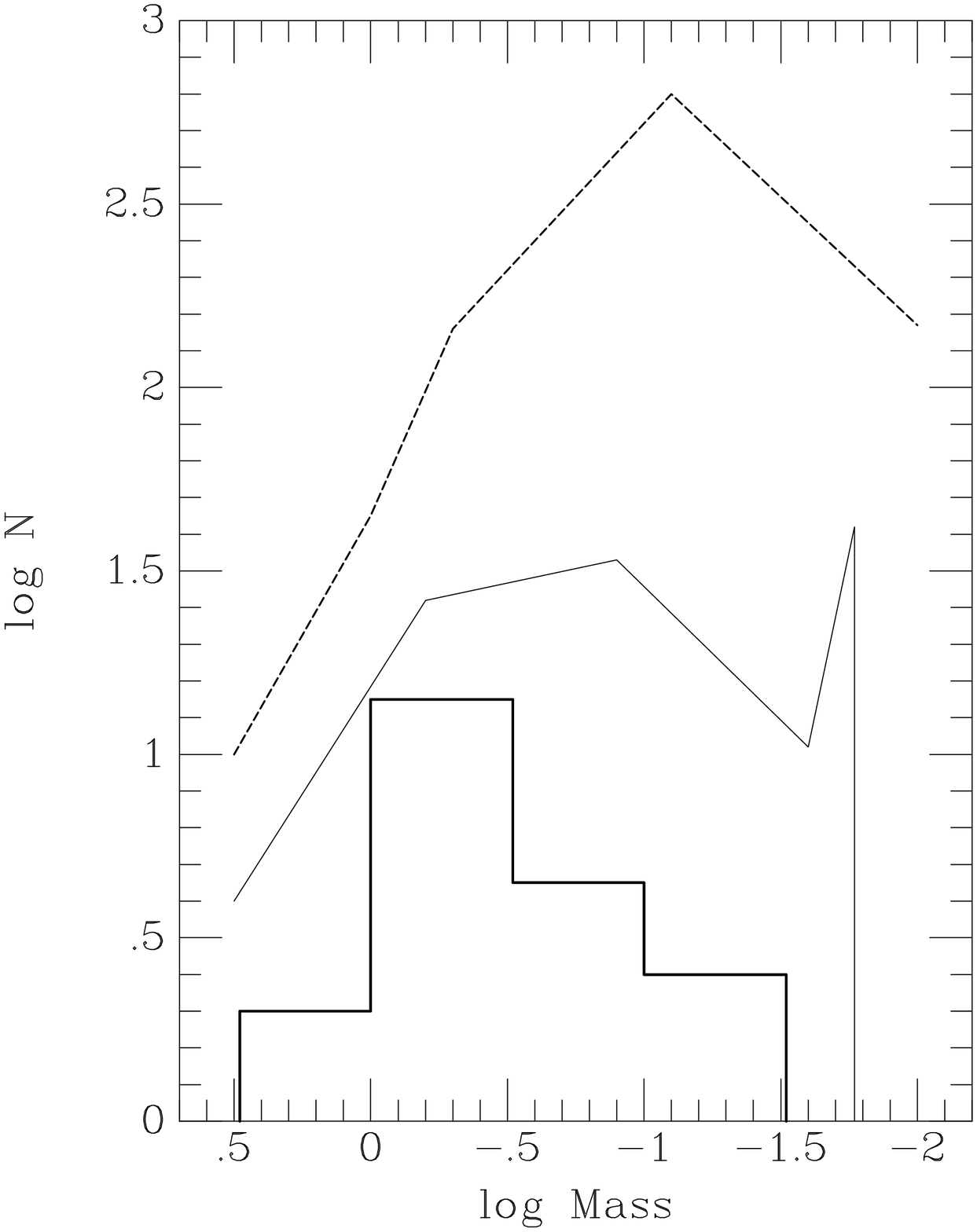}
\caption{The (a) $K$-band luminosity function and (b) Initial Mass
Function of the 27 Cha I North stars (thick histogram) compared to the
Orion Nebula Cluster (thin histogram) adjusted for distance.  The Orion
KLF and IMF are from \citet{Muench02}.  The Galactic stellar IMF from 
\citep{Kroupa01} is shown as a dashed line with arbitrary normalization.  
Chamaeleon star properties are obtained from Table \ref{oprop.tab}.
\label{LF.fig}}
\end{figure}

Having established that our sample of 27 stars is largely complete
and uncontaminated to $K \simeq 12$, we readily construct the
$K$-band luminosity based on $K$ magnitudes in Table
\ref{oprop.tab}. Figure \ref{LF.fig}a compares compares our Cha I
North KLF to that of the Orion Nebula Cluster derived by
\citet{Muench02} after scaling (as in Figure \ref{Lx-K.fig}) by
$\Delta K = 2.0$ to account for the difference in distance.  We
see that both KLFs show a similar rise from bright magnitudes to a
peak around $K \simeq 9.5$. The fall off at faint magnitudes
appears somewhat steeper in Cha I North, but the difference is not
statistically significant.

Figure \ref{LF.fig}b shows the Cha I North IMF for our sample using
masses given in Table \ref{oprop.tab}, and compares it to both the IMF
inferred for the ONC \citep{Hillenbrand00} and a general Galactic IMF
derived by \citet{Kroupa01} from a variety of cluster and field star
populations. Here, a deficiency of lower mass stars in Cha I North is
clearly present.  The Cha I North IMF peaks in the $0.3-1.0$ M$_\odot$
bin while the other IMFs peak around 0.1 M$_\odot$.  The effect is
statistically significant.  A Kolmogorov-Smirnov two-sample test
between the ONC and Chamaeleon I North IMFs indicate only a 0.3\%
probability (3$\sigma$ equivalent) they are drawn from the same
population.  Alternatively, if one considers the ratio of stars in the
(0.1-0.3):(0.3-1.0) M$_\odot$ bins, the observed ratio in Cha I North
of 5:14 has a 99\% confidence ratio $0.09-0.72$ while the Orion and
Kroupa IMFs predict a ratio around 1.5.   The difference can be erased
only if a considerable number of Cha I North stars are present that
simultaneously have $K$ below our completeness limit $\simeq 12$ and
masses above $M \simeq 0.1$ M$_\odot$; i.e., an older PMS population
superposed on the younger well-characterized population.

We thus establish a deficit of $0.1-0.3$ M$_\odot$ stars in Cha I
North compared to standard IMFs We can tentatively and qualitatively
extend this inference to the BD regime. Three previously known
objects at or below the stellar limit are detected: ISO 192, ISO
217 and Hn 13. But no additional X-ray selected population (with
the possible exception of the rapidly variable source \#42 with no
reported O/NIR counterpart) is seen. If a large number of BDs were
present, we would expect a fraction of them to have X-ray
luminosities above our sensitivity and appear in the ACIS image.
This result is consistent with reports of poor BD populations in
the Taurus-Auriga and IC 348 star forming regions \citep{Luhman00,
Briceno02, Preibisch03}. However, this argument can not be made
quantitatively until the COUP survey establishes the scatter about
the $L_t-L_{bol}$ and $L_t-K$ relationships in substellar PMS
objects.

It may be that the Cha I North cloud has an intrinsically
non-standard IMF due to its star formation process. For example,
the population of lowest mass stars may be sensitive to the
spectrum of the turbulent velocity field in the cloud
\citep{DelgadoDonate04}. But, it is critical to recall that the
deficiency of lower mass stars applies only to the $16\arcmin
\times 16\arcmin$\/ ($0.8 \times 0.8$ pc) ACIS field. We thus can
not exclude alternative hypotheses that place the lower mass stars
preferentially outside the ACIS field of view. It seems plausible
that mass segregation is present leading to a surfeit of higher
mass stars in the cloud core. This could either be a
characteristic of the primordial cluster formation process or a
later dynamical development. Evidence for primordial mass
segregation has been found both in the rich Orion Nebula Cluster
\citep{Bonnell98} and in the sparse $\eta$ Cha cluster
\citep{Lyo04}.  Dynamical segregation could also occur if lower
mass stars are preferentially born with a velocity dispersion
greater than the higher mass stars, or if they are ejected from
multiple systems due to close gravitational encounters
\citep[e.g.][]{Reipurth01, Kroupa03b}. A velocity dispersion
difference as small as $\geq 0.2$ km s$^{-1}$ is sufficient for
stars to travel outside the ACIS field in $\simeq 2$ Myr, a
typical age of the Cha I North stars.

\section{Conclusions}

The low-mass population of the Chamaeleon I cloud has been
investigated in the O/NIR bands by several groups but with
discrepant samples and differing conclusions \citep{Cambresy98,
Persi00, Comeron00, Gomez03, LopezMarti04, Luhman04}. Some are
optimistic that a significant population of low mass stars and
substellar objects are being found within the heavily contaminated
infrared samples, while others suggest BDs are deficient or that
the census is too incomplete to reach clear conclusions. We bring
to bear here a distinct and complementary method for identifying
PMS stars of all masses: high sensitivity, high resolution imagery
with the $Chandra$ X-ray Observatory. Our $Chandra$ survey with
limiting $\log L_t \simeq 27$ erg s$^{-1}$, sufficient to detect
the contemporary active Sun, should be complete to $K \simeq 12$
or $M \geq 0.1$ M$_\odot$.  Equally importantly, X-ray surveys of
this type can confidently tell when an O/NIR star is $not$ a cloud
member because of the enormous difference in $L_x/L_{bol}$ ratios
for PMS and old disk stars which frequently contaminate O/NIR
surveys.

We find no cloud members with $26.9-27.5 < \log L_t < 28.3$ erg
s$^{-1}$ ($0.5-8$ keV band), and no more than one new X-ray discovered
cloud member is present.  We furthermore confirm that several
previously suspected cloud members are probably background stars.  The
result is a nearly complete and uncontaminated sample of 27 cloud
members in the ACIS field of the Cha I North cloud.  The IMF of this
sample is significantly deficient in $0.1-0.3$ M$_\odot$ stars compared
to the IMF of the rich Orion Nebula Cluster and of the general stellar
IMF.  This supports recent observational studies that report deficits
in the BD population in low density star formation regions
\citep{Briceno02, Preibisch03}.  However, our X-ray field of view is
rather small, so we can not discriminate between a true deficiency in
low mass stars and alternative explanations such as mass segregation or
dynamical ejection of the lowest mass PMS stars\footnote{As this study 
goes to press, \citet{Stelzer04} announced results from an $XMM-Newton$
X-ray study of the Cha I South region.  Their image is larger but
several times less sensitive than our $Chandra$ ACIS image.  They
detect two bona fide brown dwarfs and several H$\alpha$ emitting
objects near the hydrogen burning mass limit. It is possible that
Cha I South does not share the same IMF as the inner region of Cha
I North we study here.}.

From the forthcoming COUP study, we are likely to learn how the
relationships between $L_t$ and $L_{bol}$, $K$ and $M$ extend into
the substellar regime.  This will permit us to derive quantitative
inferences regarding the BD population in Cha I North. At
the present time, we can say only qualitatively that this
population appears to be small as we would expect some fraction to
have X-ray luminosities exceeding $\log L_t \simeq 27$ erg
s$^{-1}$.  Three of the 27 stars in our complete sample --
ISO 192, ISO 217 and (probably) Hn 13 -- lie below the substellar
limit and, by virtue of their X-ray emission, are confirmed cloud
members.

The reliability of the X-ray census method used here rests on the
empirical relationships between $L_t$ and $K$ and the closely related
correlations between $L_t$, $L_{bol}$ and masses $M$ among PMS stars.
These relationships have been seen in nearly all well-studied PMS
stellar clusters with $Chandra$: the ONC \citep{Flaccomio03,
Feigelson03}, NGC 1333 \citep{Getman02}, IC 348 \citep{Preibisch02},
Mon R2 \citep{Kohno02} and NGC 2024 \citep{Skinner03}.  Given their
importance, it is worrisome that these relationships between magnetic
activity and bulk stellar properties are largely unexplained
\citep{Feigelson03}. But if one accepts them, these relationships
permit a translation between an observational X-ray flux limit and
limits in the KLF and IMF. Our confidence in the reliability and
completeness of X-ray surveys will increase as we improve our
characterization of these relationships and scatter about them.

Finally, we bring to attention one remarkable low mass PMS star.
ISO 192 was already known to have the heaviest absorption and
strongest NIR color excess among Cha I members with heavy
spectroscopic veiling and large photometric variability. It has
been interpreted as a substellar ($\sim 60$ M$_J$) Class I
protostar and the likely source of the Cha I North molecular
bipolar flow.  We add here another unusual characteristic: strong
magnetic activity with intrinsic X-ray luminosity around $10^{29}$
erg s$^{-1}$.

{\it Acknowledgements} ~~ Kevin Luhman (CfA) generously shared
results from unpublished Magellan spectra, A-Ran Lyo (UNSW@ADFA)
assisted with the optical spectroscopy, and Lisa Crause (UCT)
helped with the optical photometry. We thank the SAAO and MSSSO
Telescope Allocation Committees for observing time, and EDF
appreciates the University of New South Wales and Australian
Defence Force Academy for hospitality during much of this work. We
greatly benefitted from the SIMBAD and USNO databases, and made
use of data products from the Two Micron All Sky Survey which is a
joint project of the University of Massachusetts and the Infrared
Processing and Analysis Center/California Institute of Technology
funded by NASA and the NSF. This study was principally supported
by NASA's $Chandra$ Guest Observer program grant GO1-2005 (EDF,PI)
and UNSW@ADFA URSP, FRG and SRG research grants (WAL, PI).

\newpage

\begin{deluxetable}{rrrrrcccc}
\centering \tablecolumns{9} \tabletypesize{\small}
\tablewidth{0pt}

\tablecaption{$Chandra$ sources and counterparts in Cha I North \label{xsrcs.tab}}

\tablehead{\multicolumn{6}{c}{X-ray source} &&
\multicolumn{2}{c}{Stellar counterpart} \\ \cline{1-6} \cline{8-9}

\colhead{\#} & \colhead{R.A.} & \colhead{Dec.} &
\colhead{$\theta$} & \colhead{Cts} & \colhead{CHRX} &&
\colhead{Star} & \colhead{$\Delta$} \\

\colhead{(1)} & \colhead{(2)} & \colhead{(3)} & \colhead{(4)} &
\colhead{(5)} & \colhead{(6)} && \colhead{(7)} & \colhead{(8)} }

\startdata

  1 & 11 07 13.6  & -76 38 55.2 & 10.9 &   12 & \nodata && \nodata  & \nodata \\
  2 & 11 07 34.4  & -76 34 33.9 &  8.9 &   30 & \nodata && \nodata  & \nodata \\
  3 & 11 07 45.1  & -76 35 32.3 &  8.3 &   37 & \nodata && \nodata  & \nodata \\
  4 & 11 08 10.2  & -76 29 01.8 &  8.8 &   15 & \nodata && \nodata  & \nodata \\
  5 & 11 08 18.9  & -76 32 54.9 &  6.7 &   13 & \nodata && \nodata  & \nodata \\
  6 & 11 08 21.2  & -76 33 51.6 &  6.3 &   19 & \nodata && \nodata  & \nodata \\
  7 & 11 08 23.9  & -76 33 07.8 &  6.3 &    4 & \nodata && \nodata  & \nodata \\
  8 & 11 08 40.8  & -76 36 07.8 &  5.3 &  990 &    33   && CHXR 33  & 0.4     \\
  9 & 11 08 41.5  & -76 34 51.0 &  5.0 &   49 & \nodata && \nodata  & \nodata \\
 10 & 11 08 42.8  & -76 36 36.8 &  5.3 &   33 & \nodata && \nodata  & \nodata \\
 11 & 11 08 44.3  & -76 34 38.2 &  4.9 &   22 & \nodata && \nodata  & \nodata \\
 12 & 11 08 46.0  & -76 26 23.9 &  9.5 &   12 & \nodata && \nodata  & \nodata \\
 13 & 11 08 48.3  & -76 40 55.4 &  7.8 &   10 & \nodata && \nodata  & \nodata \\
 14 & 11 08 50.9  & -76 38 00.1 &  5.6 &    6 & \nodata && \nodata  & \nodata \\
 15 & 11 08 50.9  & -76 25 14.5 & 10.4 &   42 & \nodata &&   T 37   & 1.0     \\
 16 & 11 08 51.4  & -76 38 26.3 &  5.8 &   54 & \nodata &&  Field?  & 0.2     \\
 17 & 11 08 56.6  & -76 33 41.2 &  4.3 &    7 & \nodata && \nodata  & \nodata \\
 18 & 11 09 01.2  & -76 33 56.9 &  4.0 &   17 & \nodata && \nodata  & \nodata \\
 19 & 11 09 01.3  & -76 34 17.0 &  3.9 &   34 & \nodata && \nodata  & \nodata \\
 20 & 11 09 05.3  & -76 23 51.4 & 11.4 &   31 & \nodata && \nodata  & \nodata \\
 21 & 11 09 07.4  & -76 32 16.2 &  4.3 &   14 & \nodata && \nodata  & \nodata \\
 22 & 11 09 10.6  & -76 27 12.5 &  8.2 &    8 & \nodata &&  Field   & 2.1     \\
 23 & 11 09 11.8  & -76 31 37.0 &  4.5 &    8 & \nodata && \nodata  & \nodata \\
 24 & 11 09 13.5  & -76 30 23.7 &  5.3 &   23 & \nodata &&  Field   & 0.6     \\
 25 & 11 09 13.9  & -76 28 40.0 &  6.8 &  158 &    35   && CHXR 35  & 0.5     \\
 26 & 11 09 17.8  & -76 27 58.2 &  7.3 & 2348 &    37   && CHXR 37  & 0.5     \\
 27 & 11 09 18.3  & -76 30 29.5 &  5.1 &  843 &    79   && CHXR 79  & 0.5     \\
 28 & 11 09 22.3  & -76 39 03.0 &  5.1 &    6 & \nodata && \nodata  & \nodata \\
 29 & 11 09 22.6  & -76 34 32.0 &  2.6 &   26 & \nodata && C 1-6    & 0.4     \\
 30 & 11 09 22.7  & -76 28 48.7 &  6.4 &   16 & \nodata && \nodata  & \nodata \\
 31 & 11 09 27.2  & -76 41 38.7 &  7.4 &   22 & \nodata && \nodata  & \nodata \\
 32 & 11 09 27.3  & -76 34 30.1 &  2.4 &    9 & \nodata && \nodata  & \nodata \\
 33 & 11 09 28.5  & -76 33 28.2 &  2.6 &   46 & \nodata && ISO 192  & 0.1     \\
 34 & 11 09 31.1  & -76 29 01.7 &  6.1 &   16 & \nodata && \nodata  & \nodata \\
 35 & 11 09 31.5  & -76 28 22.1 &  6.7 &   16 & \nodata && \nodata  & \nodata \\
 36 & 11 09 33.3  & -76 32 52.3 &  2.7 &    6 & \nodata && \nodata  & \nodata \\
 37 & 11 09 38.4  & -76 38 47.5 &  4.5 &   10 & \nodata && \nodata  & \nodata \\
 38 & 11 09 40.1  & -76 28 39.5 &  6.2 & 2846 &    40   && CHXR 40  & 0.5     \\
 39 & 11 09 40.9  & -76 29 46.5 &  5.1 &   33 & \nodata && \nodata  & \nodata \\
 40 & 11 09 41.6  & -76 30 42.2 &  4.3 &    7 & \nodata && \nodata  & \nodata \\
 41 & 11 09 41.9  & -76 34 58.5 &  1.5 & 2328 & \nodata && C 1-25   & 0.1     \\
 42 & 11 09 45.1  & -76 30 22.2 &  4.5 &   16 & \nodata && \nodata  & \nodata \\
 43 & 11 09 46.2  & -76 34 46.4 &  1.3 &   75 & \nodata && Hn 10E   & 0.1     \\
 44 & 11 09 50.1  & -76 36 47.6 &  2.4 & 1105 &   42    &&   T 41   & 0.1     \\
 45 & 11 09 51.1  & -76 27 15.3 &  7.5 &   42 & \nodata && \nodata  & \nodata \\
 46 & 11 09 52.1  & -76 39 12.9 &  4.6 &   17 & \nodata && ISO 217  & 0.3     \\
 47 & 11 09 53.4  & -76 34 25.6 &  0.9 &  699 & \nodata &&   T 42   & 0.1     \\
 48 & 11 09 54.1  & -76 29 25.4 &  5.3 & 1189 &    41   &&   T 43   & 0.1     \\
 49 & 11 09 54.4  & -76 31 11.6 &  3.6 &   60 & \nodata && ISO 225  & 0.4     \\
 50 & 11 09 54.8  & -76 25 35.0 &  9.1 &   17 & \nodata &&  Field   & 1.8     \\
 51 & 11 09 55.0  & -76 32 40.9 &  2.1 &   37 & \nodata && C 1-2    & 0.1     \\
 52 & 11 09 56.6  & -76 27 35.7 &  7.1 &   21 & \nodata && \nodata  & \nodata \\
 53 & 11 10 00.1  & -76 34 57.9 &  0.5 & 2516 &    44   &&   T 44   & 0.1     \\
 54 & 11 10 00.7  & -76 43 29.2 &  8.8 &   25 & \nodata && \nodata  & \nodata \\
 55 & 11 10 03.7  & -76 33 29.3 &  1.2 &   62 & \nodata && Hn 11    & 0.2     \\
 56 & 11 10 04.7  & -76 35 45.2 &  1.1 & 2055 &    45   &&   T 45a  & 0.0     \\
 57 & 11 10 06.3  & -76 29 48.4 &  4.9 &   73 & \nodata && \nodata  & \nodata \\
 58 & 11 10 07.0  & -76 29 37.9 &  5.1 & 1781 &    46   &&   T 46   & 0.3     \\
 59 & 11 10 07.9  & -76 34 23.3 &  0.3 &   63 & \nodata && \nodata  & \nodata \\
 60 & 11 10 09.4  & -76 41 48.7 &  7.2 &   16 & \nodata && \nodata  & \nodata \\
 61 & 11 10 11.4  & -76 35 29.2 &  0.8 &  331 & \nodata &&  ISO 237 & 0.0     \\
 62 & 11 10 12.1  & -76 37 26.7 &  2.8 &   11 & \nodata && \nodata  & \nodata \\
 63 & 11 10 13.8  & -76 37 39.0 &  3.0 &    6 & \nodata && \nodata  & \nodata \\
 64 & 11 10 22.4  & -76 27 16.1 &  7.5 &  143 & \nodata && \nodata  & \nodata \\
 65 & 11 10 27.5  & -76 27 19.3 &  7.5 &   79 & \nodata && \nodata  & \nodata \\
 66 & 11 10 32.1  & -76 41 09.5 &  6.6 &    8 & \nodata && \nodata  & \nodata \\
 67 & 11 10 36.2  & -76 34 10.3 &  1.7 &   19 & \nodata && \nodata  & \nodata \\
 68 & 11 10 36.9  & -76 31 32.6 &  3.6 &   42 & \nodata && \nodata  & \nodata \\
 69 & 11 10 41.3  & -76 34 09.9 &  2.0 &    4 & \nodata && \nodata  & \nodata \\
 70 & 11 10 46.8  & -76 45 14.7 & 10.9 &   46 & \nodata && \nodata  & \nodata \\
 71 & 11 10 47.2  & -76 30 44.2 &  4.6 &    7 & \nodata && \nodata  & \nodata \\
 72 & 11 10 52.5  & -76 29 16.3 &  6.0 &   59 & \nodata && \nodata  & \nodata \\
 73 & 11 10 52.6  & -76 28 09.8 &  7.0 &  438 &    80   && \nodata  & \nodata \\
 74 & 11 10 53.3  & -76 34 31.9 &  2.6 &   73 &    82   &&   T 48   & 0.0     \\
 75 & 11 10 55.9  & -76 45 32.8 & 11.2 &  106 & \nodata && Hn 13    & 0.4     \\
 76 & 11 10 58.3  & -76 40 03.9 &  6.1 &   13 & \nodata && \nodata  & \nodata \\
 77 & 11 11 01.8  & -76 32 36.4 &  3.7 &   12 & \nodata && \nodata  & \nodata \\
 78 & 11 11 04.0  & -76 39 03.2 &  5.4 &   12 & \nodata && \nodata  & \nodata \\
 79 & 11 11 05.1  & -76 29 41.4 &  6.0 &   19 & \nodata &&  Field   & 0.4     \\
 80 & 11 11 06.4  & -76 32 39.8 &  3.9 &   90 & \nodata && \nodata  & \nodata \\
 81 & 11 11 09.8  & -76 37 13.0 &  4.4 &   15 & \nodata && \nodata  & \nodata \\
 82 & 11 11 10.1  & -76 36 27.0 &  4.0 &    7 & \nodata && \nodata  & \nodata \\
 83 & 11 11 10.6  & -76 32 33.1 &  4.2 &    4 & \nodata && \nodata  & \nodata \\
 84 & 11 11 16.9  & -76 28 57.9 &  7.0 &   12 & \nodata &&  Field   & 0.7     \\
 85 & 11 11 19.7  & -76 34 49.0 &  4.2 &   17 & \nodata && \nodata  & \nodata \\
 86 & 11 11 20.0  & -76 38 25.7 &  5.6 &   40 & \nodata &&  Field   & 0.8     \\
 87 & 11 11 21.6  & -76 35 12.0 &  4.3 &    5 & \nodata && \nodata  & \nodata \\
 88 & 11 11 23.6  & -76 33 55.1 &  4.4 &    7 & \nodata &&  Field   & 2.0     \\
 89 & 11 11 24.2  & -76 37 43.1 &  5.4 &  322 &    83   && \nodata  & \nodata \\
 90 & 11 11 25.9  & -76 36 06.2 &  4.7 &   70 & \nodata && \nodata  & \nodata \\
 91 & 11 11 34.7  & -76 36 21.2 &  5.3 & 1768 &    48   && CHXR 48  & 0.2     \\
 92 & 11 11 38.5  & -76 30 40.2 &  6.6 &   37 & \nodata && \nodata  & \nodata \\
 93 & 11 11 40.8  & -76 38 59.9 &  6.9 &   17 & \nodata && \nodata  & \nodata \\
 94 & 11 11 41.3  & -76 36 39.0 &  5.7 &    8 & \nodata && \nodata  & \nodata \\
 95 & 11 11 44.9  & -76 40 16.8 &  7.9 &   24 & \nodata && \nodata  & \nodata \\
 96 & 11 11 47.2  & -76 29 33.1 &  7.7 &   10 & \nodata && \nodata  & \nodata \\
 97 & 11 11 49.1  & -76 30 29.8 &  7.2 &  158 & \nodata && \nodata  & \nodata \\
 98 & 11 12 00.1  & -76 31 28.2 &  7.3 &   10 & \nodata &&  Field   & 0.2     \\
 99 & 11 12 00.2  & -76 34 30.0 &  6.5 &   21 & \nodata &&  Field   & 0.4     \\
100 & 11 12 03.2  & -76 37 03.3 &  7.1 &  113 &    84   && CHXR 84  & 0.3     \\
101 & 11 12 04.5  & -76 33 43.2 &  6.8 &   60 & \nodata && \nodata  & \nodata \\
102 & 11 12 05.4  & -76 31 37.6 &  7.5 &  235 & \nodata && \nodata  & \nodata \\
103 & 11 12 09.7  & -76 34 35.4 &  7.1 &   60 &    85   &&   T 50   & 1.3     \\
104 & 11 12 19.6  & -76 31 43.8 &  8.2 &   63 & \nodata && \nodata  & \nodata \\
105\tablenotemark{a} & 11 12 23.2  & -76 37 04.3 &  8.2 &   50 &    50   &&   T 51   & 4.8     \\
106 & 11 12 26.3  & -76 35 30.2 &  8.1 &   14 & \nodata && \nodata  & \nodata \\
107 & 11 12 46.9  & -76 32 04.2 &  9.6 &   27 & \nodata && \nodata  & \nodata \\

\enddata

\tablecomments{Table columns: \\
1. Running number for this paper.  \\
2-3. Position from the ACIS image aligned to 2MASS sources.
Generally, positions are accurate to $<0.3$\arcsec\/ for off-axis
angle $\theta<5$\arcmin and $Cts > 100$.  Positional accuracy
degrades to $\sim 1$\arcsec\/ for fainter on-axis sources, and to
$2-3$\arcsec\/ for sources far off-axis ($\theta > 10$\arcmin).\\
4. Off-axis angle in arcmin.   \\
5. $Cts$ are the $0.5-8$ keV events in the 95\% enclosed-energy polygon
after subtraction of background and rounded to the nearest integer
obtained from {\it acis\_extract}.\\
6. CHXR is the $ROSAT$ source number from Feigelson et al.\ (1993). \\
7. Name from the cloud membership list by Carpenter et al. (2002).
`Field' indicates previously unremarked stellar
counterparts. The remainder of the X-ray sources are probably extragalactic. \\
8. Offset in arcsec between the Chandra and 2MASS positions. }

\tablenotetext{a}{Source \#105 lies at the edge of the ACIS
detector where, due to the broad telescope point spread function
and satellite dithering, many of the source photons may be lost.
This can result in an underestimate of the source luminosity and
an apparent positional offset of the centroid.}

\end{deluxetable}

\newpage

\begin{deluxetable}{rcrrccccrrr}
\centering \tablecolumns{11} \tabletypesize{\small}
\tablewidth{0pt}

\tablecaption{X-ray properties of $Chandra$ Cha I North stellar sources
\label{xprop.tab}}

\tablehead{

\colhead{\#} & \colhead{Name} & \colhead{$C_x$} & \colhead{$B_x$}
& \colhead{Var} & \colhead{$\log N_H$} & \colhead{$kT_1$} &
\colhead{$kT_2$} & \colhead{$\log L_s$} & \colhead{$\log L_t$} &
\colhead{$\log L_c$}  \\

\colhead{(1)} & \colhead{(2)} & \colhead{(3)} & \colhead{(4)} &
\colhead{(5)} & \colhead{(6)} & \colhead{(7)} & \colhead{(8)} &
\colhead{(9)} & \colhead{(10)} & \colhead{(11)} }

\startdata

  8 & CHXR 33 &  993 &  3 & c & 21.8 & 0.7$\pm 0.3$ & \nodata & 29.3 & 29.4 & 30.0    \\
 15 & T 37    &   68 & 24 & a &  0   & 2.0$\pm 1.3$ & \nodata & 27.9 & 28.1 & 28.1    \\
 16 & Field?  &   57 &  3 & a & 22.1 & 1.3$\pm 0.4$ & \nodata & 27.9 & 28.3 & 28.7    \\
 22 & Field   &   15 &  7 & a &  0:  & 2:           & \nodata & 27.1 & 27.4 & 27.4    \\
 24 & Field   &   25 &  2 & a & 22.3 & 0.3          & \nodata & 27.4 & 27.4 & \nodata \\
 25 & CHXR 35 &  164 &  6 & a & 21.8 & 0.2          & 1.0     & 28.6 & 28.6 & 30.1    \\
 26 & CHXR 37 & 2357 &  9 & a & 21.7 & 0.4          & 2.2     & 29.6 & 29.8 & 30.4    \\
 27 & CHXR 79 &  845 &  2 & a & 21.8 & 0.6$\pm 0.1$ & \nodata & 28.9 & 29.7 & 29.8    \\
 29 & C 1-6   &   27 &  1 & c & 22.4 & 2.2$\pm 1.3$ & \nodata & 27.3 & 28.3 & 28.8    \\
 33 & ISO 192 &   47 &  1 & a & 23:  & 2:           & \nodata & 27.1 & 28.6 & \tablenotemark{a}~~~ \\
 38 & CHXR 40 & 2853 &  7 & a &  0   & 0.8          & 2.5     & 29.8 & 29.9 & 29.9    \\
 41 & C 1-25  & 2329 &  1 & c & 22.2 & 10:          & \nodata & 29.2 & 30.3 & 30.4    \\
 43 & Hn 10E  &   76 &  1 & a & 22.0 & 0.9$\pm 0.2$ & \nodata & 28.1 & 28.2 & 28.9    \\
 44 & T 41    & 1106 &  1 & c & 21.3 & 0.8          & 3.2     & 29.2 & 29.5 & 29.7    \\
 46 & ISO 217 &   17 &  2 & c & 22.4 & 5:           & \nodata & 27.1 & 28.3 & 28.5    \\
 47 & T 42    &  699 &  0 & b & 21.8 & 0.7          & 4.9     & 29.2 & 29.7 & 30.0    \\
 48 & T 43    & 1192 &  3 & c & 21.8 & 0.4          & 2.5     & 29.6 & 29.6 & 30.1    \\
 49 & ISO 225 &   61 &  1 & b & 22.7 & 10:          & \nodata & 26.8 & 28.9 & 29.1    \\
 50 & Field   &   25 &  8 & a & 21.9 & 1.3$\pm 2.1$ & \nodata & 27.5 & 27.8 & 28.3    \\
 51 & C 1-2   &   38 &  1 & b & 22.4 & 3.2$\pm 2.1$ & \nodata & 27.4 & 28.5 & 28.8    \\
 53 & T 44    & 2517 &  1 & c & 21.8 & 0.6          & 2.8     & 29.8 & 30.2 & 30.5    \\
 55 & Hn 11   &   63 &  1 & b & 21.9 & 2.1$\pm 0.8$ & \nodata & 27.9 & 28.4 & 28.7    \\
 56 & T 45a   & 2056 &  1 & b & 21.6 & 0.8          & 2.4     & 29.5 & 29.8 & 30.1    \\
 58 & T 46    & 1784 &  3 & c & 21.1 & 0.6          & 2.4     & 29.6 & 29.7 & 29.8    \\
 61 & ISO 237 &  332 &  1 & a & 22.0 & 0.7          & 4.2     & 28.8 & 29.2 & 29.7    \\
 74 & T 48    &   74 &  1 & a & 0    & 0.8$\pm 0.2$ & \nodata & 28.2 & 28.4 & 28.4    \\
 75 & Hn 13   &  130 & 24 & c & 22.4 & 1.4$\pm 0.4$ & \nodata & 28.0 & 28.8 & 29.6    \\
 79 & Field   &   23 &  4 & b & 21:  & 0.5$\pm 0.6$ & \nodata & 27.6 & 27.6 & 28.6    \\
 84 & Field   &   18 &  6 & a & 0    & 1.1$\pm 0.4$ & \nodata & 27.3 & 27.3 & 27.3    \\
 86 & Field   &   44 &  4 & c & 0    & 10:          & \nodata & 27.8 & 28.3 & 28.3    \\
 88 & Field   &    8 &  1 & a & 0    & 1:           & \nodata & 27.1 & 27.2 & 27.2    \\
 91 & CHXR 48 & 1771 &  3 & c & 21.4 & 0.4          & 1.7     & 29.6 & 29.7 & 30.0    \\
 98 & Field   &   15 &  5 & a & 0    & 0.8$\pm 0.2$ & \nodata & 27.6 & 27.6 & 27.6    \\
 99 & Field   &   25 &  4 & a & 0    & 2.1$\pm 1.1$ & \nodata & 27.6 & 27.8 & 27.8    \\
100 & CHXR 84 &  119 &  6 & a & 0    & 1.0$\pm 0.1$ & \nodata & 28.4 & 28.5 & 28.5    \\
103 & T 50    &   65 &  5 & a & 22.3 & 0.9$\pm 0.4$ & \nodata & 28.0 & 28.3 & 29.4    \\
105\tablenotemark{b} & T 51    &   53 &  3 & a & 21.8 & 0.8$\pm 0.2$ & \nodata & 28.6 & 28.7 & 29.3    \\

\enddata

\tablecomments{The X-ray luminosities assume the star lies at a
distance of 160 pc.  The Field stars discussed in \S \ref{new.sec}
lie on the far side of the cloud at unknown distances, so their
X-ray luminosities are greater than the tabulated values.}

\tablenotetext{a}{See text, \S \ref{known.sec}.}

\tablenotetext{b}{See note to Table \ref{xsrcs.tab}.}

\end{deluxetable}

\newpage

\begin{deluxetable}{rcccrrrrrrrccrlr}
\centering \rotate \tablecolumns{16} \tabletypesize{\footnotesize}
\tablewidth{0pt}

\tablecaption{Optical-infrared properties of $Chandra$ Cha I North
stellar sources \label{oprop.tab}}

\tablehead{

\colhead{\#} & \colhead{Star} & \colhead{Prop Flag} &
\colhead{Class} &
\multicolumn{7}{c}{Photometry} && \multicolumn{4}{c}{HR Diagram}
\\ \cline{5-11} \cline{13-16}

 & & & & \colhead{R} & \colhead{i/I} & \colhead{J} & \colhead{H} &
\colhead{K} & \colhead{m(6.7)} & \colhead{m(14.3)} & &
\colhead{SpTy} & \colhead{L$_{bol}$} & \colhead{Mass} & \colhead{Age} \\

\colhead{(1)} & \colhead{(2)} & \colhead{(3)} & \colhead{(4)} &
\colhead{(5)} & \colhead{(6)} & \colhead{(7)} & \colhead{(8)} &
\colhead{(9)} & \colhead{(10)} & \colhead{(11)} & & \colhead{(12)}
& \colhead{(13)} & \colhead{(14)} & \colhead{(15)} }

\startdata

  8 &  CHXR 33  &  0YYY00 & WTT     &  14.5  & 12.59   & 10.56 &  9.66 &    9.28 &  8.73   & \nodata && M0      &  ~0.42  &  0.7    &  3      \\
 15 &    T 37   &  0Y00Y0 & CTT     & \nodata& 14.60   & 12.45 & 11.73 &   11.30 & 10.1~   & \nodata && M5.25   &  ~0.04? &  0.1    &  8      \\
 16 &  Field?   & \nodata & \nodata &  17.4  & 16.39   & 14.58 & 13.63 &   13.33 & \nodata & \nodata && \nodata & \nodata & \nodata & \nodata \\
 22 &  Field    & \nodata & \nodata &  17.0  & 16.13   & 14.65 & 13.86 &   13.51 & \nodata & \nodata && \nodata & \nodata & \nodata & \nodata \\
 24 &  Field    & \nodata & \nodata &  19.6  & 17.88   & 15.01 & 13.64 &   13.05 & \nodata & \nodata && \nodata & \nodata & \nodata & \nodata \\
 25 &  CHXR 35  &  0Y0Y00 & CTT     &  16.1  & 13.91   & 11.85 & 11.21 &   10.87 & \nodata & \nodata && M4.75   &  ~0.072 &  0.2    &  3      \\
 26 &  CHXR 37  &  0YYY00 & WTT     &  13.2  & 11.72   & 10.00 &  9.04 &    8.70 &  8.47   &  8.3~   && K7      &  ~1.0   &  0.8    &  1      \\
 27 &  CHXR 79  &  YY0YY0 & CTT     &  16.8  & 14.99   & 11.66 & 10.12 &    9.07 &  7.12   &  5.55   && M1.25   &  ~0.47  &  0.6    &  2      \\
 29 &  C 1-6    &  Y000Y0 & CTT     &  19.5  & 16.79   & 12.60 & 10.35 &    8.67 &  5.84   &  3.73   && M1.25   &  ~0.77  &  0.6    &  1      \\
 33 &  ISO 192  &  Y000Y0 & I/BD    & \nodata& \nodata & 16.07 & 13.27 &   11.04 &  6.40   &  2.83   && \nodata & \nodata & \nodata & \nodata \\
 38 &  CHXR 40  &  0YYY00 & WTT     &  12.9  & 11.66   & 10.07 &  9.23 &    8.96 &  8.73   & \nodata && M1.25   &  ~0.69  &  0.6    &  1      \\
 41 &  C 1-25   &  Y000Y0 & WTT     & \nodata& \nodata & 13.80 & 11.42 &   10.00 &  7.05   &  4.91   && \nodata & \nodata & \nodata & \nodata \\
 43 &  Hn 10E   &  0Y00YY & CTT     &  16.7  & 14.70   & 11.95 & 10.74 &   10.05 &  7.7    &  5.8    && M3.25   &  ~0.18  &  0.25   &  3      \\
 44 &    T 41   &  000YYY & AB/WTT  &   8.8  &  8.82   &  7.64 &  7.35 &    7.35 &  2.84   &  0.93   && B9      &  61.    &  2.7    &  1      \\
 46 &  ISO 217  &  0000Y0 & BD?     &  19.4  & 16.63   & 13.53 & 12.54 &   11.82 &  9.8~   & \nodata && M6.25   &  ~0.028 &  0.07   &  3      \\
 47 &    T 42   &  YY00YY & CTT     &  17.0  & 11.79   &  9.47 &  7.79 &    6.46 &  3.58   &  1.57   && K5      &  ~3.5   &  1.0:   & $<$1    \\
 48 &    T 43   &  YY0YY0 & CTT     &  15.8  & 14.03   & 11.30 & 10.00 &    9.25 &  7.66   &  5.91   && M2      &  ~0.46  &  0.55   &  3      \\
 49 &  ISO 225  &  Y000Y0 & CTT     &  18.4  & 17.23   & 15.05 & 13.80 &   13.14 &  9.3~   &  7.6~   && M1.75   &  ~0.013 &  0.5:   & \nodata \\
 50 &  Field    & \nodata & \nodata &  15.4  & 14.70   & 13.60 & 12.95 &   12.71 & \nodata & \nodata && \nodata & \nodata & \nodata & \nodata \\
 51 &  C 1-2    &  Y000YY & CTT     & \nodata& 17.85   & 13.78 & 11.30 &    9.67 &  6.34   &  4.34   && \nodata & \nodata & \nodata & \nodata \\
 53 &    T 44   &  YY0YY0 & CTT     &  12.6  & 10.95   &  8.71 &  7.21 &    6.08 &  3.62   &  1.64   && K5      &  ~6.1   &  1.0:   & $<$1    \\
 55 &  Hn 11    &  0Y00Y0 & CTT     &  15.9  & 14.70   & 11.77 & 10.26 &    9.44 &  7.25   &  5.66   && K8      &  ~0.61  &  0.75   &  2      \\
 56 &    T 45a  &  Y00YYY & WTT     &  15.2  & 12.38   & 10.57 &  9.37 &    9.24 &  8.1~   &  6.6~   && M0      &  ~0.45  &  0.7    &  3      \\
 58 &    T 46   &  Y00YYY & CTT     &  13.1  & 11.62   &  9.91 &  8.96 &    8.45 &  6.83   &  5.20   && K8      &  ~0.75  &  0.75   &  2      \\
 61 &   ISO 237 &  0000Y0 & CTT     &  16.1  & 13.75   & 10.93 &  9.44 &    8.62 &  6.63   &  4.75   && K5.5    &  ~1.3   &  0.9    &  1      \\
 74 &    T 48   &  YY0YY0 & CTT     &  14.1  & 13.07   & 11.26 & 10.45 &   10.04 &  8.02   &  6.29   && M1      &  ~0.15  &  0.6    & 10      \\
 75 &  Hn 13    &  0Y00Y0 & CTT     &  16.3  & 13.95   & 11.16 & 10.42 &    9.91 &  8.56   &  7.09   && M5.75   &  ~0.16  &  0.08   &  6      \\
 79 &  Field    & \nodata & \nodata &  17.5  & 16.06   & 14.37 & 13.57 &   13.23 & \nodata & \nodata && \nodata & \nodata & \nodata & \nodata \\
 84 &  Field    & \nodata & \nodata &  15.5  & \nodata & 14.40 & 14.04 &   13.83 & \nodata & \nodata && \nodata & \nodata & \nodata & \nodata \\
 86 &  Field    & \nodata & \nodata &  20.5  & \nodata & 16.8: & 15.7: &$>$15.0~ & \nodata & \nodata && \nodata & \nodata & \nodata & \nodata \\
 88 &  Field    & \nodata & \nodata &  18.1  & \nodata & 16.04 & 15.16 &   15.0: & \nodata & \nodata && \nodata & \nodata & \nodata & \nodata \\
 91 &  CHXR 48  &  YYYY00 & WTT     &  13.4  & 12.35   & 10.86 & 10.08 &    9.80 &  8.02   &  6.29   && M2.5    &  ~0.25  &  0.4    &  3      \\
 98 &  Field    & \nodata & \nodata &  16.0  & \nodata & 14.55 & 14.13 &   14.08 & \nodata & \nodata && \nodata & \nodata & \nodata & \nodata \\
 99 &  Field    & \nodata & \nodata &  15.5  & \nodata & 14.19 & 13.66 &   13.54 & \nodata & \nodata && \nodata & \nodata & \nodata & \nodata \\
100 &  CHXR 84  &  0Y0Y00 & WTT     &  16.2  & 13.97   & 11.77 & 11.11 &   10.78 & \nodata & \nodata && M5.5    &  ~0.082 &  0.12   &  1      \\
103 &    T 50   &  YY0Y00 & CTT     &  15.4  & 13.41   & 10.96 & 10.18 &    9.84 & \nodata & \nodata && M5      &  ~0.19  &  0.2    & $<$1    \\
105 &    T 51   &  0Y0Y0Y & WTT     &   9.9  & 10.22   &  9.28 &  8.52 &    8.00 & \nodata & \nodata && K3.5    &  ~1.1   &  1.2    &  5      \\

\enddata

\tablecomments{3. Six character flag indicating the properties of that star from
Carpenter et al.\ (2002): Variable in optical or K; H$\alpha$
emission; Li absorption; X-ray source (prior to the present
study); infrared excess; far-infrared source. \\
4. WTT = weak-lined T Tauri star, CTT = classical T Tauri star,
I = imbedded, BD = brown dwarf, AB = Herbig AeBe star. \\
5-11. $R$ magnitudes from the USNOB-1.0 catalog, $i$ or $I$ magnitudes
from the Second DENIS data release or our SAAO $I$ survey, $JHK$ magnitudes
from the 2MASS survey, and 6.7-$\mu$m and 14.3-$\mu$m magnitudes from the
ISO survey. \\
12-15. Properties for cloud members from Luhman (2004).}

\end{deluxetable}

\clearpage
\newpage

\clearpage
\newpage

\clearpage
\newpage

\clearpage
\newpage

\clearpage
\newpage

\clearpage
\newpage

\clearpage
\newpage

\end{document}